\documentclass[english]{article}
\usepackage[T1]{fontenc}
\usepackage[latin9]{inputenc}
\usepackage{geometry}
\usepackage{color}
\usepackage{babel}
\usepackage{refstyle}
\usepackage{float}
\usepackage{textcomp}
\usepackage{amsmath}
\usepackage{graphicx}
\usepackage{setspace}
\usepackage[unicode=true,pdfusetitle,
 bookmarks=true,bookmarksnumbered=false,bookmarksopen=false,
 breaklinks=false,pdfborder={0 0 1},backref=false,colorlinks=false]
 {hyperref}
\usepackage{breakurl}

\makeatletter

\RS@ifundefined{subsecref}
  {\newref{subsec}{name = \RSsectxt}}
  {}
\RS@ifundefined{thmref}
  {\def\RSthmtxt{theorem~}\newref{thm}{name = \RSthmtxt}}
  {}
\RS@ifundefined{lemref}
  {\def\RSlemtxt{lemma~}\newref{lem}{name = \RSlemtxt}}
  {}

\newref{fig}{%
    name      = \RSFigtxt,  
    names     = \RSFigstxt,
    Name      = \RSFigtxt,
    Names     = \RSFigstxt,
    rngtxt    = \RSrngtxt,
    lsttwotxt = \RSlsttwotxt,
    lsttxt    = \RSlsttxt}

\makeatother

\begin{document}
\begin{doublespace}
\begin{center}
\textbf{\textcolor{black}{\Large{}A Tale of Two Consequences:}}{\Large\par}
\par\end{center}

\begin{center}
\textbf{\Large{}Intended and Unintended Outcomes of the Japan TOPIX
Tick Size Changes}{\Large\par}
\par\end{center}

\begin{center}
\textbf{Ravi Kashyap }
\par\end{center}

\begin{center}
\textbf{SolBridge International School of Business / City University
of Hong Kong}
\par\end{center}

\begin{center}
\begin{center}
\today
\par\end{center}
\par\end{center}

\begin{center}
Japan; Venue; Analysis; Tick; Size; Change; Exchange; Execution; Uncertainty;
Costs; Trading
\par\end{center}

\begin{center}
\par\end{center}

\begin{center}
\textbf{\textcolor{blue}{\href{https://doi.org/10.3905/jot.2015.10.4.051}{Edited Version: Kashyap, R. (2015). A Tale of Two Consequences. The Journal of Trading, 10(4), 51-95. }}}\tableofcontents{}
\par\end{center}
\end{doublespace}
\begin{doublespace}

\section{Abstract }
\end{doublespace}

\begin{doublespace}
We look at the effect of the tick size changes on the TOPIX 100 index
names made by the Tokyo Stock Exchange on Jan-14-2014 and Jul-22-2014.
The intended consequence of the change is price improvement and shorter
time to execution. We look at security level metrics that include
the spread, trading volume, number of trades and the size of trades
to establish whether this goal is accomplished. An unintended effect
might be the reduction in execution sizes, which would then mean that
institutions with large orders would have greater difficulty in sourcing
liquidity. We look at a sample of real orders to see if the execution
costs have gone up across the orders since the implementation of this
change. 

We study the mechanisms that affect how securities are traded on an
exchange, before delving into the specifics of the TSE tick size events.
Some of the topics we explore are: The Venue Menu and How to Increase
Revenue; To Automate or Not to Automate; Microstructure under the
Microscope; The Price of Connections to High (and Faraway) Places;
Speed Thrills but Kills; Pick a Size for the Perfect Tick; TSE Tick
Size Experiments, Then and Now; Sergey Bubka and the Regulators; Bird\textquoteright s
Eye View; Deep Dive; Possibilities for a Deeper Dive; Does Tick Size
Matter? Tick Size Does Matter!
\end{doublespace}
\begin{doublespace}

\section{The Venue Menu and How to Increase Revenue}
\end{doublespace}

\begin{doublespace}
The more that shoppers shop, the more shops there will be and the
more the shops will try to woo the shoppers. Similar has been the
effect of increasing financialization across the globe. Greater levels
of trading, by both retail and institutional investors, have resulted
in more exchanges springing up and offering more products that can
be traded. The longer the menu of venues, more the competition among
them, and this naturally requires attempts at trying to attract and
retain customers, by a venue to increase its revenue. If people are
willing to pay (or bid) more than what is asked (or offered), then
perhaps, we would not have specialized venues to trade, a small price
to pay for, let us just say, peace on Earth. Forgetting about Utopia
- but keeping in mind that conceivably, the Bid-Offer spread can be
a barometer to a civilization's progress, till it becomes irrelevant,
indicating that a society has transcended beyond mere material matters
of accumulating and allocating wealth - a brief and worthy digression
would be to look at how exchanges have evolved and what factors drive
the future development of trading venues. 

An exchange, as the word implies, is the process during which people
give and take things of similar value. At a place where this transfer
happens, also an exchange, shares or holdings can be liquidated and
hence the primary mission of an exchange is to provide liquidity.
For the rest of the discussion, we ignore the exchange of OTC (Over
The Counter) securities, which are traded wherever, whenever and however
one can trade them; but we leave the reader with the analogy that
if Exchange Trading is similar to collecting tolls on a road; OTC
Trading is like highway robbery. 

As with most historical matters, there is no agreement on when and
where the first stock exchange was started. There seems to be some
consensus that the first exchanges were started to finance East India
companies that provided investment for merchants that sailed on the
high seas to conduct business with various countries in Asia. In addition
to raising capital, these trading arenas, offered a means for the
transfer and sharing of risk. The effect of this simple movement of
securities between owners, ripples across, gets magnified and affects
the entire economy. See, (Michie 2001), for an excellent exposition
on the questions exchanges faced centuries ago and how they are similar
to the issues that are cropping up today, with a focus on the London
Stock Exchange (LSE). Conceptually, the trading venues of today, still
perform the same duties, since all of finance, through time, has involved
three simple outcomes \textendash{} \textquotedblleft Buy, Sell or
Hold\textquotedblright . The complications are mainly to get to these
results.
\end{doublespace}
\begin{doublespace}

\subsection{To Automate or Not To Automate}
\end{doublespace}

\begin{doublespace}
Alongside the progression of exclusive trading locations, a parallel
development has been the increased use of automation and technology
in the buying and selling of securities. This has removed the traditional
concept of a brick and mortar building where specialists or jobbers
(the market makers on the NYSE - New York Stock Exchange - and LSE)
acted as the counter-parties for brokers, who were undertaking orders
on behalf of the end investors. Since the advent of the National Association
of Securities Dealers Automated Quotations (NASDAQ), the first electronic
stock market, many more electronic trading networks have proliferated
and virtual trading has been adopted even by the remaining physical
exchanges to a great degree. 

There is no dearth of evidence that the rules of trading affect the
profitability of various trading strategies. Venkataraman (2001) compares
securities on the NYSE (floor-based trading structure with human intermediaries,
specialists and floor brokers) and the Paris Bourse (automated limit
order trading structure). He finds that execution costs might be higher
on automated venues even after controlling for differences in adverse
selection, relative tick size and economic attributes. A trade occurs
when an aggressive trader submits a market order and demands liquidity,
hence the rules on a venue are designed to attract demanders of liquidity,
and nudge liquidity providers to display their orders. Displaying
limit orders involves risks. First, the counter-parties could be better
informed and liquidity providers could get picked off. Hence, they
would like the trading system to allow them to trade selectively with
counter-parties of their choice. Second, they risk being front-run
by other traders with an increase in the market impact of their orders.
Hence, large traders want to hide their orders and expose them only
to traders who are most likely to trade with them. This means fully
automated exchanges, which anecdotally seems to be the way ahead,
need to take special care to formulate rules to help liquidity providers
better control the risks of order exposure. 
\end{doublespace}
\begin{doublespace}

\subsection{Microstructure under the Microscope}
\end{doublespace}

\begin{doublespace}
Market microstructure is the investigation of the process and protocols
that govern the exchange of assets with the objective of reducing
frictions that can impede the transfer. In financial markets, where
there is an abundance of recorded information, this translates to
the study of the dynamic relationships between observed variables,
such as price, volume and spread, and hidden constituents, such as
transaction costs and volatility, that hold sway over the efficient
functioning of the system. Madhavan (2000) provides a comprehensive
survey of the theoretical and experimental literature relating to:
price formation, the dynamic process by which prices come to impound
information; market structure and design, the relation between price
formation and trading rules; and transparency, the ability of market
participants to observe information about the trading process.

Differences in microstructure, such as tick size, spread, trade depth
and clustering of prices (clustering is the tendency for prices to
fall on a subset of available prices), could be due to differences
in market structure such as whether a given market is a dealer market
or an auction market. Auction markets are order driven, where buy
orders seek the lowest available prices and sell orders seek the highest
available prices. This process is called the price discovery process
because it reveals the prices that best match buyers to sellers. Dealer
markets are quote-driven because prices are set only by dealer quotes
in the market. Huang and Stoll (2001) study securities on the LSE
and American Depository Receipts (ADRs) on the NYSE. The key feature
differentiating the two market structures is the treatment of public
limit orders. In an auction market, limit orders are displayed and
may trade against incoming market orders. In a dealer market, limit
orders are held by each dealer, are not displayed, and can only be
traded against the dealer\textquoteright s quote. Spreads could be
higher in dealer markets since they are set by dealers. A minimum
tick is necessary in an auction market to encourage liquidity provision
by limit orders and by dealers. Without a minimum tick (or a minimum
trade size), a limit order can cheaply step ahead of another limit
order or a dealer quote. If there is no minimum tick, it is easy to
avoid time priority. Dealer markets do not require time priority across
dealers and they have less need for a minimum tick. However, each
dealer quotes in depth even in the absence of a tick rule because
he wishes to maintain a reputation for liquidity or because dealer
markets set standards as to depth. Quote clustering is highly correlated
with spreads and with the stock characteristics that determine spreads.
If a market has higher spreads it has greater clustering. Trade clustering
is lesser relative to quote clustering in dealer markets indicating
that negotiations with dealers can be successful even though negotiation
for better prices by customers takes place off the screen whereas
negotiation in an auction market takes place on the screen via limit
order placement. In auction markets, limit orders break up quote clustering
as they seek to gain priority and trades cluster to a similar degree
as quotes. Higher spreads are accompanied by greater depth. Trade
sizes are larger consistent with the large depth, but the difference
in trade size is not as great as the difference in depths.

While a number of variables can be observed on an exchange, the primary
lever available for adjustment is the spread. (Roll 1984 and Stoll
1989) connect stock price changes to the bid-offer spread. The spread
is determined due to order processing costs, adverse information or
inventory holdings costs. A key distinction to be aware of is between
the quoted spread and the realized spread. The quoted spread is the
difference between the ask price quoted by a dealer and the bid price
quoted by a dealer at a point in time. The realized bid-ask spread
is the average difference between the price at which a dealer sells
at one point in time and the price at which a dealer buys at an earlier
point in time. The quoted spread is related to characteristics of
securities such as the volume of trading, the stock price, the number
of market makers, the volatility, and other factors. If the spread
reflects only order processing costs, the bid and the offer always
straddle the true price. The dealer covers costs by buying at the
bid and selling at the offer (on average). Sequences of purchases
at the bid price are ultimately offset by sequences of sales at the
ask price. In this case, the realized spread and quoted spread are
the same. An implication of both the inventory cost model and the
adverse information cost model is that the realized spread earned
by a dealer is less than the spread quoted by the dealer (empirical
studies show this to be the case; references mentioned earlier in
the paragraph have more details). Under the inventory cost model,
this is because the dealer lowers both bid and ask prices after a
dealer purchase in order to induce dealer sales and inhibit additional
dealer purchases and raises both bid and ask prices after a sale in
order to induce dealer purchases and inhibit dealer sales. The net
effect of bid and ask price changes are such that future transactions
that will equilibrate inventory. New prices are set such that the
dealer is indifferent between a transaction at the bid price and a
transaction at the ask price. Under the adverse information cost model,
bid and ask prices are changed in a similar way to reflect the information
conveyed by transactions. After a sale to the dealer, bid and ask
prices are lowered because a transaction conveys information that
the expected equilibrium price of the security is lower. Transactions
convey information under the assumption that some traders are better
informed than others.
\end{doublespace}
\begin{doublespace}

\subsection{The Price of Connections to High (and Faraway) Places}
\end{doublespace}

\begin{doublespace}
Technology coupled with globalization is causing financial markets
to be linked. While the global financial connection is mostly implicit
at this stage, explicit links across exchanges are being added (See
Kashyap 2015b on the Hong Kong Shanghai Connect). Caldarelli, Marsili,
and Zhang (1997) show that even a simple model of an exchange, operating
as a completely closed system with no external influences, where the
participants trade with the sole purpose of increasing their capital
after observing the price history, can produce rich and complex fluctuations
in prices. In our numerical explorations, we return to this premise
and restrict our observations to measurements that can be directly
gathered on an exchange. Structuring the study in this way abstracts
away from the subjective decision regarding what external variables
can influence a system. This benchmark scenario, which is a simplified
platform to glean illuminative lessons, differs from a realistic setting
with regards to the changes in demand from an exchange participant,
which can be influenced by external factors as observed by the participant
himself or from the order flow he receives from investors that are
not exchange members, who could be acting due to other extraneous
forces. Karolyi and Stulz (1996) explore factors that affect cross-country
stock return correlations using dollar-denominated returns of U.S.
and Japanese shares trading in the U.S. They find that U.S. macroeconomic
announcements, shocks to the Yen/Dollar foreign exchange rate and
Treasury bill returns, and industry effects have no measurable influence
on U.S. and Japanese return correlations or the co-movements between
U.S. and Japanese share returns. However, large shocks to broad-based
market indices (Nikkei Stock Average and Standard and Poor's 500 Stock
Index) positively impact both the magnitude and persistence of the
return correlations. Possibly, movements in prices are the biggest
contributors to further price movements.

Adding technology enabled buying and selling, which can support more
trades, to the ability to buy and sell in distant lands, accelerates
the transfer of securities, causing prices to be displaced back and
forth from any equilibrium (rather pseudo equilibrium) which causes
more transactions to happen affecting the entire cycle of investment
management (Kashyap 2014a). Advocates of setting daily limits on how
much the price of a security can change over the course of a trading
day believe that such measures can decrease stock price volatility,
counter overreaction, and do not interfere with trading activity.
Daily price limit critics claim that price limits cause higher volatility
levels on subsequent days (volatility spillover hypothesis), prevent
prices from efficiently reaching their equilibrium level (delayed
price discovery hypothesis), and interfere with trading due to limitations
imposed by price limits (trading interference hypothesis). Kim and
Rhee (1997) study the effect of daily price limits on the Tokyo Stock
Exchange (TSE) and despite small sample size issues find evidence
supporting all three hypotheses, suggesting that price limits may
be ineffective. 

This constant jumping of prices and the associated transfer of capital
gives rise to a highly specialized work force that seeks to plug every
source of inefficiency and profit from it (Kashyap 2015a). As funds
flow from the benefactor to the beneficiary, chunks of it are taken
by players along the financial food chain. The touch point during
this transfer of capital, the exchange, has to seen as reducing any
potential losses. But intermediaries and their actions can possibly
aid the overall process of liquidation, making it quicker and efficient;
it also brings more trading volumes, a main source of income for the
exchanges. There is also an indirect channel as more participants
and more volume begets lower spreads, which lowers execution costs,
which induces more volume, which then generates more profits. Easley
and O'Hara (2010) demonstrate the potential benefits to exchanges,
investors and firms from reducing ambiguity over how markets work
or asset prices are formed. Uncertainty can cause some traders to
be overly influenced by \textquotedblleft worst case\textquotedblright{}
outcomes, even when these outcomes have little objective possibility
of occurring. This, in turn, can cause such naive investors to opt
not to participate in markets, a result detrimental to both markets
and the economy alike. Microstructure features (which in their study
refers to listing thresholds; monitoring to ensure fair and non-manipulative
trading; and operational oversight of clearing and settlement to insure
that a trader who buys stock actually receives it and that one who
sells stock actually delivers it) can be used to reduce this ambiguity,
and thereby induce greater participation in markets. A side effect
of excess liquidity might be investor passivity and fragmentation
of stockholdings (Bhide 1993), since investors can exit stocks they
don't like easily and their holdings are not big enough to be a voice
for better corporate governance.
\end{doublespace}
\begin{doublespace}

\subsection{Speed Thrills but Kills}
\end{doublespace}

\begin{doublespace}
The technical arms race will give rise to a situation where participants
focus their efforts on trading faster once certain kinds of new information
is received since being first would mean the difference between profits
and losses (or perhaps, just the difference between profits and lesser
profits, which can sometimes seem equally worse in an atmosphere where
milliseconds matter). Ye, Yao \& Gai (2012) confirm the old adage,
speed thrills but kills. They find evidence that increasing the speed
of trading from the microsecond level to the nanosecond level, lead
to dramatic increases in message flow. The increases in message flow
are due largely to increases in order cancellations without any real
increases to actual trading volume. Spread does not decrease following
increase in speed; market efficiency, in terms of price formation,
does not improve; market depth decreases and short-term volatility
increases, probably as a consequence of more cancellations. A fight
for speed increases high-frequency order cancellation but not real
high-frequency order execution. Increased cancellation generates more
noise to the message flow. Low-frequency traders then subsidize high-frequency
traders because only executed trades are charged a fee. The exchanges
continually make costly system enhancements to accommodate higher
message flow, but these enhancements facilitate further order cancellations,
not increases in trading volume. Investment in high frequency trading
with sub-millisecond accuracy may provide a private benefit to traders
without consummate social benefit; therefore, there may be an over-investment
in speed.

The point that warrants further consideration is whether High Frequency
Trading (HFT) is leading to benefits by either directly providing
additional liquidity or indirectly via the spawning of numerous technological
innovations in computer networking hardware, software or other items
used to facilitate HFT, which can then be beneficial to other sectors.
An example that concerns high speeds is from the car racing industry,
which comes up with innovations that produce faster and safer cars.
Many of these innovations slip into the mainstream automobile industry
over time. Completely restricting any endeavor is not ideal since
it hard to know where the next life changing idea might spring up;
but regulating the dangerous or unfavorable ones is prudent. Learning
further from this example, we do not see race-cars cruising down our
town streets (though, not something to rule out entirely in the not
too distant future); they hustle around in an exclusive arena, indicating
that perhaps we need a similar mandate for the HFT industry.
\end{doublespace}
\begin{doublespace}

\subsection{Pick a Size for the Perfect Tick}
\end{doublespace}

\begin{doublespace}
From a theoretical point of view (Harris 1994), the tick size is the
lower bound of the bid-ask spread. We can then expect that a reduction
in the tick size would decrease the quoted spread. Nevertheless, the
reduction in the spread could also decrease order exposure because
liquidity provision is less profitable and more risky. As a consequence,
the quoted depth could also decline. Goldstein and Kavajecz (2000)
and Lipson and Jones (2001) explore the effects of the tick size reduction
on the NYSE and find that while decrease in tick sizes might improve
the liquidity for small size orders, institutional traders were worse
off because they had to bear an increase in trading costs following
the decline in depth throughout the entire order book. Bessembinder
(2003) reports evidence regarding trade execution costs and market
quality after the 2001 decimilization on NYSE and NASDAQ, which includes
narrower average quoted, effective, and realized bid-ask spreads on
both markets, lower volatility on both markets, and the absence of
systematic reversals of quote changes on either market, indicating
that market quality has indeed been improved, while admitting that
a complete assessment of the impact of decimalization on market quality
will also require access to proprietary data on institutional trading
programs, in order to assess whether trading costs for large institutions
have also declined.

Bourghelle and Declerck (2004) look at the consequences of a change
in the tick size at the Paris Bourse, where there was a both a decrease
and increase in tick size on different groups of securities and hence
offered an opportunity to simultaneously explore the issues involved.
Decreased tick sizes induced a decrease in depth at the quotes. However,
in contrast with results obtained for US markets, (reflecting differences
in market design between European and US exchanges), this change neither
generated a change in the bid-ask spread nor a reduction in liquidity
provision for large trades. Limit order submission inside the best
quotes (on the best quotes) increases significantly, and investors
use more hidden quantity orders to reduce exposure. Stocks that experienced
an increase in the tick size did not have altered spreads, but it
increased the depth at the best quotes, showing evidence of a larger
display of liquidity. Limit order submission inside the best quotes
significantly decreases, and investors use fewer hidden quantity orders.
By increasing the per-share rent, a larger relative tick makes liquidity
supply more profitable and probably attracts new limit order traders
in the market. To attract liquidity demanders, designers of trading
systems have to stimulate investors to fully display their orders.
A relatively coarse pricing grid does not always result in excessively
large spreads, but enhances quoted depth, encourages liquidity providers
to expose their trading interest and stimulate investors to quote
the competitive spread.

Ahn, Cao and Choe (1998) examine the impact of decimalization in Canada
and find a significant reduction in the spread and quotation depth
on the Toronto Stock Exchange (ToSE) and a significant reduction in
the spread on NASDAQ for ToSE stocks indicating that NASDAQ dealers
might not operate as efficiently as perfect competition warrants and
could quote narrower spreads without any rule change on the NASDAQ.
However, the decimalization does not affect the spread on the NYSE
and American Stock Exchange (AMEX) for ToSE cross-listed stocks. The
most important finding is that despite an economically significant
reduction in the spread on the ToSE, orders for the cross-listed stocks
do not migrate from U.S. markets to the ToSE. This result contrasts
with the ToSE\textquoteright s objective to attract order flows from
the U.S. markets and to increase the market share of the ToSE in cross-listed
stocks. The savings in transaction costs on the ToSE are not sufficient
to offset the benefits of trading (which include the ease of trading
and superior execution of blocks) on the NYSE and AMEX. The practice
of payments for order flow has existed between Canadian brokers and
U.S. dealers for years. Given the restriction that a Canadian broker
cannot accept payments for the purchased order from a Canadian dealer
but can accept payments from a U.S. dealer, there is little incentive
to direct the order to the TSE for execution, even though the TSE
offers lower trading costs. Chung and Van Ness (2001) look at the
intraday effect on spreads due to the Order Handling Rules (OHR, which
included quote depth and tick size reductions) implemented on the
NASDAQ in 1997. They find that the tick-size reduction led to a significant
decline in spreads with the magnitude of the decline being largest
(smallest) during the last (first) hour of trading and to a significant
decrease in quoted depths with the magnitude of the decline being
smallest during the first hour of trading. Bessembinder (1999) finds
that executions costs on the NASDAQ remain higher compared to the
NYSE even after the OHR was put in place, though the cross-market
differential has decreased steadily over time. Some explanations for
the difference could be that: one, NASDAQ securities have different
economics characteristics (such as greater return volatility or smaller
investor base); two, NASDAQ's quote driven dealer structure could
be less efficient than the order drive NYSE structure; three, or the
NASDAQ facilitates a form of collusion that can keep spreads higher.
A point worth noting is the preferencing arragements whereby orders
are routed by brokers to dealer based on preexisting agreements rather
than to the market maker displaying the best quote might have been
responsible for a lack of competition on the NASDAQ. Bessembinder
(2000) examines changes in trade execution costs and market quality
for a set of NASDAQ listed firms whose tick size changed as their
share prices passed through \$10. Though there was apparently no written
rule, the convention on Nasdaq during 1995 was to use tick sizes of
1/8 dollar for bid quotations at or above \$10 per share and 1/32
dollar for bid quotations below \$10 per share. The empirical results
indicate that spreads decreased and there was no evidence of a reduction
in liquidity. 

Bollen and Busse (2006) measure changes in mutual fund trading costs
following two reductions in the tick size of U.S. equity markets:
the switch from eighths to sixteenths and the subsequent switch to
decimals. They estimate trading costs by comparing a mutual fund\textquoteright s
daily returns to the daily returns of a synthetic benchmark portfolio
that matches the fund\textquoteright s holdings but has zero trading
costs by construction. Smaller tick sizes lower depth, thereby penalizing
institutional investors. Large institutional orders are sensitive
to market depth for at least two reasons. First, filling a large order
may take several days and multiple transactions; hence a large order
likely suffers price concessions as market depth is consumed. Second,
information leakage may move prices adversely as the institutional
investor attempts to fill the order. Investors who trade small quantities
of individual equities benefit from the tighter spreads following
the switch to decimal pricing, and are largely unaffected by any decline
in depth. 

These results underscore the view that market structure has a significant
effect on trading costs and hence the design of optimal trade and
quote dissemination protocols coupled with proper regulatory oversight
of the investment process are essential for investor welfare and market
quality. Of all the weapons in the regulatory arsenal, it seems, changes
that can influence the price formation process without directly setting
price levels, hold the greatest power. One takeaway from these studies
is that regulators may be well advised to avoid reducing tick size
if they want to attract liquidity providers, and if order exposure
is profitable for a market.
\end{doublespace}
\begin{doublespace}

\subsection{TSE Tick Size Experiments, Then and Now}
\end{doublespace}

\begin{doublespace}
The TSE has tried its hand earlier at tick size changes when it introduced
a change in its minimum tick sizes on April 13, 1998. The TSE is one
of the largest limit order markets using a tick size that is a step
function of share price. The reduction in tick size therefore depends
on price ranges. Ahn, Cai, Chan and Hamao (2007) investigate the liquidity
and market quality of the stocks affected by this change. They find
that the quoted spread and the effective spread declined significantly.
Reductions in spread are greater for firms with greater tick size
reductions, greater trading activity, and higher monopoly rent proportion
in the bid-ask spread component. There is an increase in the quote
revision (relative to the number of trades), suggesting there is more
price competition among limit order traders in providing liquidity.
Although investors are more aggressive in posting quotes, there is
no definite evidence of an increase in trading volume reflecting a
decrease in depth provided to the market. 

The current change has a three phase implementation over a period
of close to two years.
\end{doublespace}
\begin{itemize}
\begin{doublespace}
\item Phase 1 was a pilot phase that went live from 14th January 2014. It
covered stocks from TOPIX 100 index and reduced tick sizes only for
stocks with quote price above �3000. Due to the high concentration
of stocks with lower market price in TOPIX, this pilot phase had an
impact on only 39 stocks. For simplicity, we consider the affected
securities based on the prices as of the ex-date. 
\item Phase 2 commenced on 22nd July 2014 over the same universe of stocks,
i.e. TOPIX 100. It introduced decimal yen tick sizes for stocks trading
below �5000. This phase had an impact on 80 stocks. For simplicity,
we consider the affected securities based on the prices as of the
ex-date.
\item Phase 3, expected in September 2015, will be based on the tick structure
of Phase 2. However, TSE will announce the final tick sizes and list
of target stocks after evaluating the impact of previous 2 phases.
\end{doublespace}
\end{itemize}
\begin{doublespace}
\begin{figure}[H]
\includegraphics[width=17cm,height=10cm]{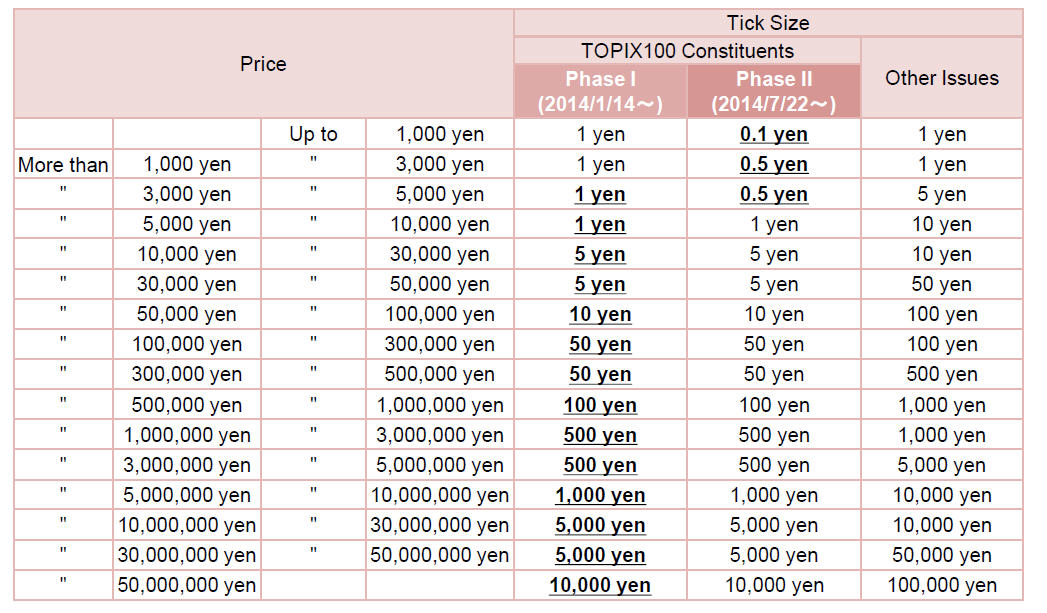}

\caption{Tick Size Change Schedule}
\end{figure}

\end{doublespace}
\begin{doublespace}

\subsection{Sergey Bubka and the Regulators}
\end{doublespace}

\begin{doublespace}
Any attempt at regulatory change is best exemplified by the story
of Sergey Bubka (End-note \ref{enu:Sergey-Nazarovich-Bubka}), the
Russian pole vault jumper, who broke the world record 35 times. Attempts
at regulatory change can be compared to taking the bar higher. In
this case, the intended effect of the change is price improvement
and shorter time to execution. We look at security level metrics that
include the spread, trading volume, number of trades and the size
of trades to establish whether this goal is accomplished. Despite
all the uncertainty (Kashyap 2014b), we can be certain of one thing,
that certain market participants will find some way over the intended
consequences, prompting another round of rule revisions, or raising
the bar, if you will. In this circumstance, an unintended effect caused
by the reactions of participants to the new rules, might be the reduction
in execution sizes, which would then mean that institutions with bigger
orders would have greater difficulty in sourcing liquidity. We look
at a sample of real orders to see if the executions costs have gone
up across larger orders since the implementation of this change. So
far, we have talked about the unknowns (or unintended consequences)
that we know about (or can anticipate). What about the unknowns that
we don't know about (or cannot even imagine). The only thing, we know
about these unknown unknowns are that, there must be a lot of them,
hence the need for us to be eternally vigilant, compelling all attempts
at risk management to make sure that the unexpected, even if it does
happen, is contained in the harm it can cause, while being cognizant
that this is easier said than done; a topic best saved for another
time.
\end{doublespace}
\begin{doublespace}

\section{Bird's Eye View}
\end{doublespace}

\begin{figure}[H]
\includegraphics{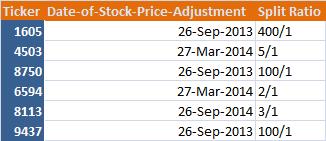}

\caption{Stock Split Ratios During Study Time Period}
\label{Stock Split Ratios}
\end{figure}

\begin{doublespace}
Before doing an in-depth study over a large sample of data, we perform
a cursory check around the days surrounding the two ex-dates to see
if we can spot any hints of change. We did three groups of high level
comparisons to ensure that the observations are not restricted to
the altered dynamics of trading on any single day. The key metrics
we used in this analysis are the average spread, measured in yen per
share and the average execution size measured in number of shares.
We adjust the price and spread according to the split ratio for securities
that had stock splits during the time period of our data sample (Figure
\ref{Stock Split Ratios}).

1. We compared the metrics on the day of the change to the metrics
on the immediately previous trading day; i.e. between Jan 14th and
Jan 10th for the first phase of the change; and between Jul 22nd and
Jul 18th for the second phase of the change (Jan 13th and Jul 21st
were public holidays in Japan).

2. We compared the metrics one and two days after the change for the
first phase and second phase respectively with the metrics exactly
a week before this day; i.e. between Jan 15th and Jan 8th; and Jul
24th and Jul 17th.

3. We compared the metrics between the earliest date in this limited
sample and the last day we have observed after the change; i.e. between
Jan 16th and Jan 6th for the first phase of the change; and between
Jul 28th and Jul 14th for the second phase of the change.
\end{doublespace}
\begin{doublespace}

\subsection{Bird's Eye View Results}
\end{doublespace}

\begin{doublespace}
As expected, after the change to decrease the tick size on the TOPIX
100 index names on Jul 22 2104 and Jan 14 2104, the average spreads
have decreased consistently, across the names that had prices in the
range which would be affected by the new rules. On the days immediately
following the change, the spreads are down for almost all the securities
affected by the change. This also fits in with the fact that the minimum
tick size after the change is 0.1 yen across some of the names. Another
observation due to this change is probably the unexpected change in
the average execution size on these names. While this change is not
as significant and consistent as the spread changes, the initial results
do indicate a trend towards decreased execution sizes.

For the first phase, the average spreads have decreased across 100\%,
100\% and 100\% of the affected names over the three sets of comparisons
that we did (It is across 63\%, 72\% and 84\% over the full set of
names and the change is across 96\% of the names when we consider
only the names where the spread was on average in excess of 1 yen
earlier) and for the second phase the spreads have decreased across
100\%, 100\% and 96\% of the affected names over the three sets of
comparisons (It is across 100\%, 90\% and 86\% of the full set of
names and the magnitude of the change is considerably smaller as compared
to the change that was done in Jan 2014). 

\begin{figure}[H]
\includegraphics{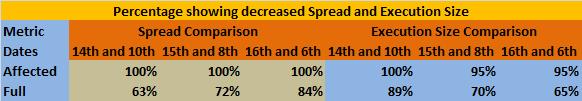}

\caption{Bird's Eye View Comparison Summary for Jan 14, 2014}
\label{BEV- Summary I}
\end{figure}

\begin{figure}[H]
\includegraphics{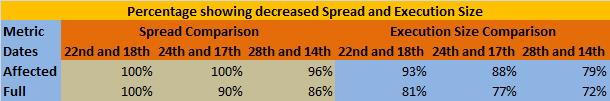}

\caption{Bird's Eye View Comparison Summary for Jul 22, 2014}
\label{BEV- Summary II}
\end{figure}

\begin{figure}[H]
\includegraphics[width=18cm,height=23.8cm]{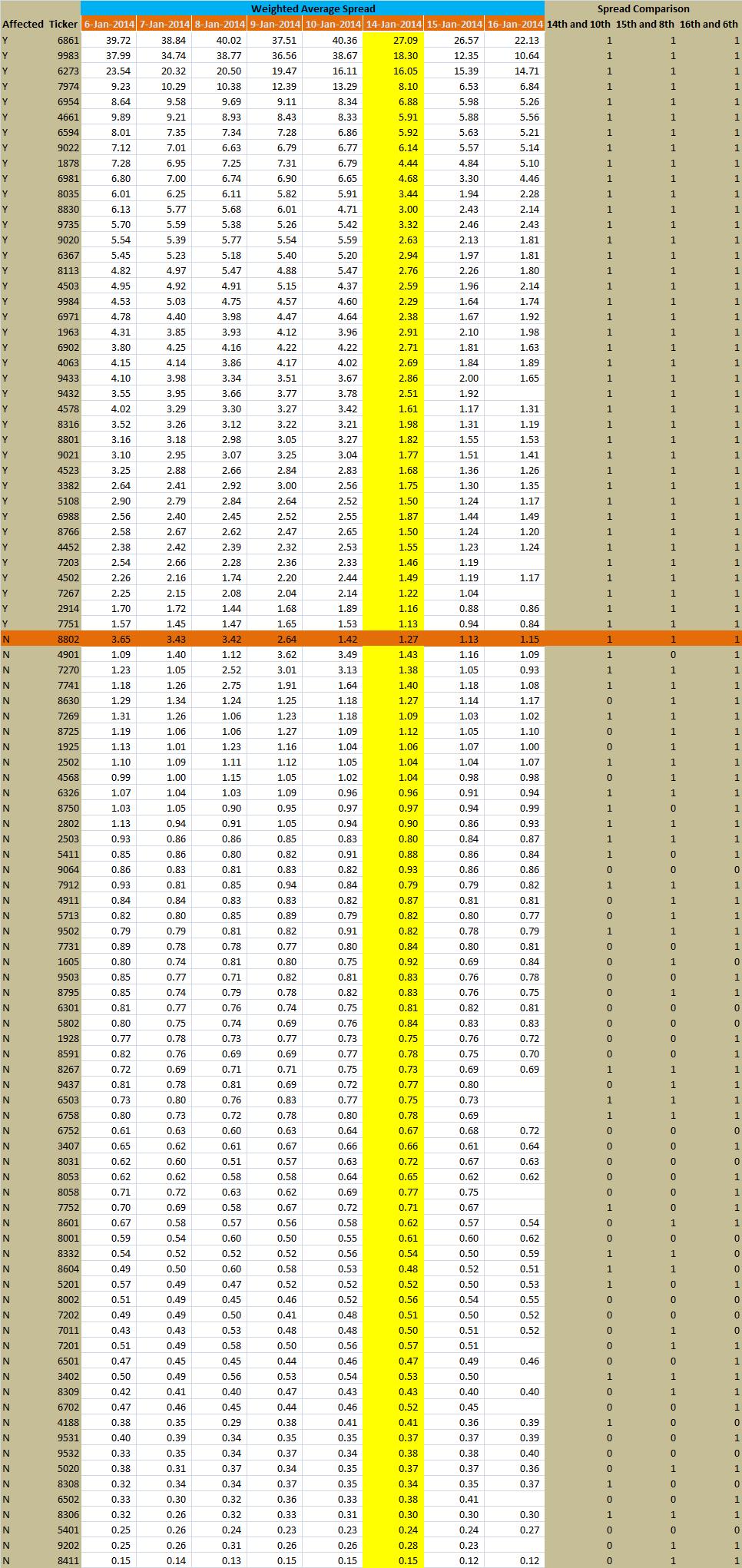}

\caption{Bird's Eye View Spread Comparison Detail for Jan 14, 2014}
\label{BEV- Details I}
\end{figure}

\begin{figure}[H]
\includegraphics[width=18cm,height=23.8cm]{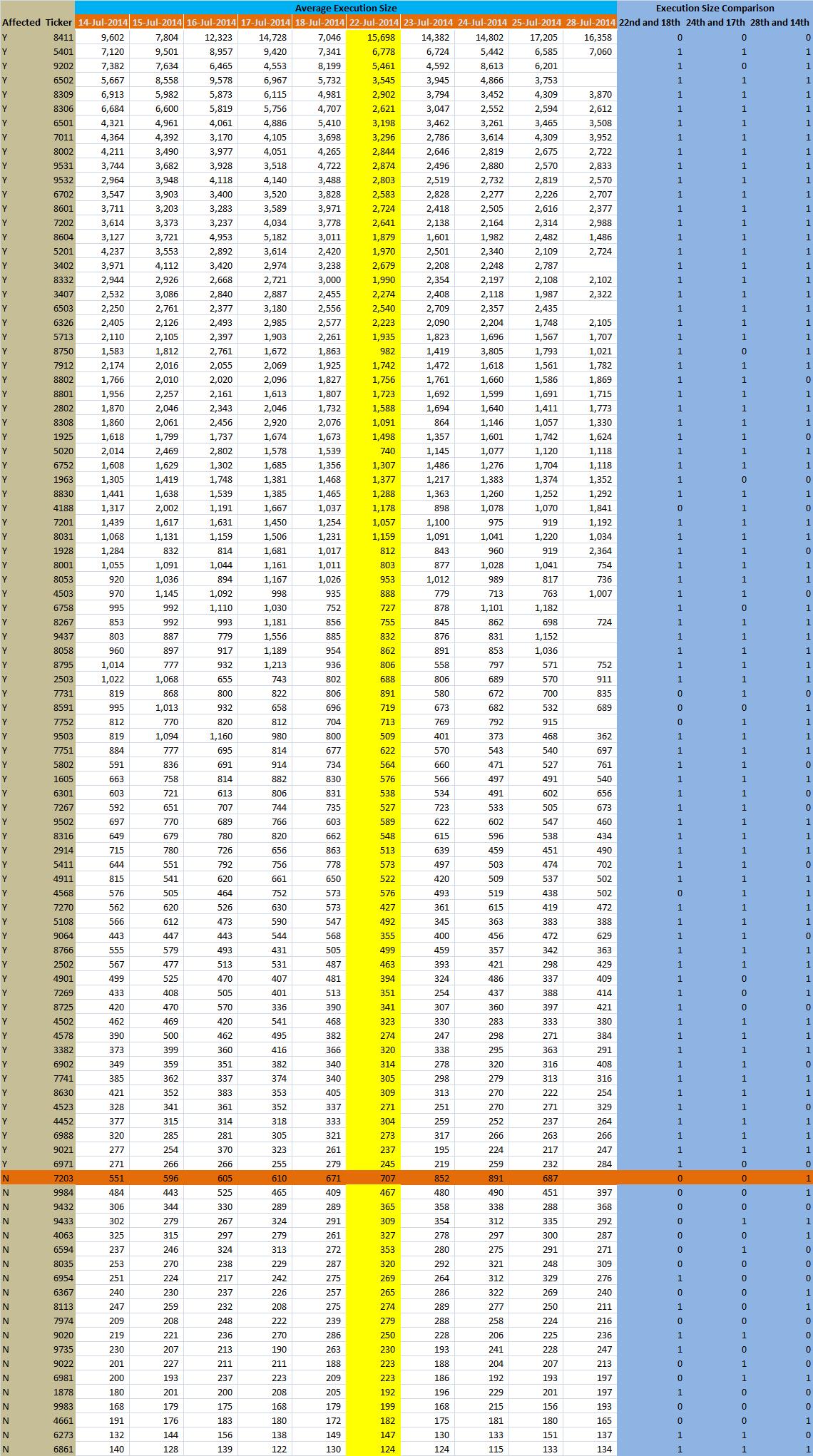}

\caption{Bird's Eye View Execution Size Comparison Detail for Jul 22, 2014}
\label{BEV- Details II}
\end{figure}

For the first phase, the average execution size has decreased across
100\%, 95\% and 95\% of the affected names over the three sets of
comparisons that we did (It is across 89\%, 70\% and 65\% over the
full set of names and there seems to be no threshold over which the
decreased size is entirely consistent, we need additional observations
for this metric) and for the second phase the average execution size
has decreased across 93\%, 88\% and 79\% of the affected names over
the three sets of comparisons (It is across 81\%, 77\% and 72\% of
the full set of names). The summary of the results (Figures \ref{BEV- Summary I}
and \ref{BEV- Summary II}) and the details (Figures \ref{BEV- Details I}
to \ref{BEV- Details II}) across individual securities are given
in section \ref{sec:Appendix---I}, Appendix - I.
\end{doublespace}
\begin{doublespace}

\section{Deep Dive}
\end{doublespace}

\begin{doublespace}
A fly through of the data across the two ex-dates reveals that a detailed
analysis would indeed be a worthy endeavor. As part of this deep dive,
first, we perform stationarity checks on prices, spreads, trade sizes
and volumes and also consider whether these variables have become
more volatile since the changes. We supplement these with statistical
tests and evaluate properties that can establish trends regarding
whether the variables are either increasing or decreasing, after the
two event dates. While this is interesting information, a question
of paramount importance is, ``what is the effect of these changes
on the trading costs across different size orders''? For this we
consider trading costs on a sample of close to 250,000 real orders
starting six months before the first event and ending six months after
the second event. Before we tackle the crucial conundrum of trading
costs, we review some basics regarding the measurement of transaction
costs.
\end{doublespace}
\begin{doublespace}

\subsection{Methodological Fundamentals}
\end{doublespace}

\begin{doublespace}
The unique aspect of our approach to trading costs is a method of
splitting the overall move of the security price during the duration
of an order into two components (Collins and Fabozzi 1991; Treynor
1994; Yegerman and Gillula 2014). One component gives the costs of
trading that arise from the decision process that went into executing
that particular order, as captured by the price moves caused by the
executions that comprise that order. The other component gives the
costs of trading that arise due to the decision process of all the
other market participants during the time this particular order was
being filled. This second component is inferred, since it is not possible
to calculate it directly (at least with the present state of technology
and publicly accessible data) and it is the difference between the
overall trading costs and the first component, which is the trading
cost of that order alone. The first and the second component arise
due to competing forces, one from the actions of a particular participant
and the other from the actions of everyone else that would be looking
to fulfill similar objectives. Naturally, it follows that each particular
participant can only influence to a greater degree the cost that arises
from his actions as compared to the actions of others, over which
he has lesser influence, but an understanding of the second component,
can help him plan and alter his actions, to counter any adversity
that might arise from the latter. Any good trader would do this intuitively
as an optimization process, that would minimize costs over two variables
direct impact and timing, the output of which recommends either slowing
down or speeding up his executions. With this measure, traders now
actually have a quantitative indicator to fine tune their decision
process. When we decompose the costs, it would be helpful to try and
understand how the two sub costs could vary as a proportion of the
total. The volatility in these two components, which would arise from
different sources (market conditions), would require different responses
and hence would affect the optimization problem mentioned above invoking
different sorts of handling and based on the situation, traders would
know which cost would be the more unpredictable one and hence focus
their efforts on minimizing the costs arising from that component.
Another popular way to decompose trading costs is into temporary and
permanent impact {[}See Almgren and Chriss (2001); Almgren (2003);
and Almgren, Thum, Hauptmann and Li (2005){]}. While the theory behind
this approach is extremely elegant and considers both linear and nonlinear
functions of the variables for estimating the impact, a practical
way to compute it requires measuring the price a certain interval
after the order. This interval is ambiguous and could lead to lower
accuracy while using this measure. 

We now introduce some terminology used throughout the discussion.
\end{doublespace}
\begin{enumerate}
\begin{doublespace}
\item Total Slippage - The overall price move on the security during the
order duration. This is also a proxy for the implementation shortfall
(Perold 1988; and Treynor 1981). It is worth mentioning that there
are many similar metrics used by various practitioners and this concept
gets used in situations for which it is not the best suited (Yegerman
and Gillula 2014). While the usefulness of the Implementation Shortfall,
or slippage, as a measure to understand the price shortfalls that
can arise between constructing a portfolio and while implementing
it, is not to be debated, slippage need to be supplemented with more
granular metrics when used in situations where the effectiveness of
algorithms or the availability of liquidity need to be gauged. 
\item Market Impact (MI) - The price moves caused by the executions that
comprise the order under consideration. In short, the MI is a proxy
for the impact on the price from the liquidity demands of an order.
This metric is generally negative or zero since in most cases, the
best impact we can have is usually no impact. 
\item Market Timing - The price moves that happen due to the combined effect
of all the other market participants during the order duration. 
\item Market Impact Estimate (MIE) - This is the estimate of the Market
Impact, explained in point two above, based on recent market conditions.
The MIE calculation is the result of a simulation which considers
the number of executions required to fill an order and the price moves
encountered while filling this order, depending on the market microstructure
as captured by the trading volume and the price probability distribution,
over the past few days. See Kashyap 2015b for a dynamic programming
approach to minimize the Market Impact under various formulations
of the law of motion of prices. This simulation can be controlled
with certain parameters that dictate the liquidity demanded on the
order, the style of trading, order duration, market conditions as
reflected by start of trading and end of trading times. In short,
the MIE is an estimated proxy for the impact on the price from the
liquidity demands of an order. Such an approach holds the philosophical
viewpoint that making smaller predictions and considering their combined
effect would result in lesser variance as opposed to making a large
prediction; estimations done over a day as compared to estimations
over a month, say. A geometrical intuition would be that fitting more
lines (or curves) over a set of points would reduce the overall error
as compared to fitting lesser number of lines (or curves) over the
same set of points. When combining the results of predictions, of
course, we have to be mindful of the errors of errors, which can get
compounded and lead the results astray, and hence, empirical tests
need to be done to verify the suitability of such a technique for
the particular situation.
\item All these variables are measured in basis points to facilitate ease
of comparison and aggregation across different groups. It is possible
to measure these in cents per share and also in dollar value or other
currency terms. 
\item The following equations, expressed in pseudo mathematical terms to
facilitate easier understanding, govern the relationships between
the variables mentioned above.
\end{doublespace}
\end{enumerate}
\begin{doublespace}

\subsection{Technical Aside }
\end{doublespace}

\begin{doublespace}
Total Slippage = Market Impact + Market Timing 

\{Total Price Slippage = Your Price Impact + Price Impact From Everyone
Else (Price Drift)\} 

Market Impact Estimate = Market Impact Prediction = f (Execution Size,
Liquidity Demand) 

Execution Size = g(Execution Parameters, Market Conditions) 

Liquidity Demand = h(Execution Parameters, Market Conditions) 

Execution Parameters <->vector comprising (Order Size, Security, Side,
Trading Style, Timing Decisions) 

Market Conditions <-> vector comprising (Price Movement, Volume Changes,
Information Set)

Here, f, g, h are functions. We could impose concavity conditions
on these functions, but arguably, similar results are obtained by
assuming no such restrictions and fitting linear or non-linear regression
coefficients, which could be non-concave or even discontinuous allowing
for jumps in prices and volumes. The specific functional forms used
could vary across different groups of securities or even across individual
securities or even across different time periods for the same security.
The crucial aspect of any such estimation is the comparison with the
costs on real orders, as outlined earlier. Simpler models are generally
more helpful in interpreting the results and for updating the model
parameters. Hamilton {[}1994{]} and Gujarati {[}1995{]} are classic
texts on econometrics methods and time series analysis that accentuate
the need for parsimonious models. 

The Auxiliary Information Set could be anything under our Sun or even
from under other heavenly objects. A useful variable to include would
be the blood pressure and heart rate time series of a representative
group of security traders.
\end{doublespace}
\begin{doublespace}

\subsection{Implementation Shortfall Refresher}
\end{doublespace}

\begin{doublespace}
As a brush up, the total slippage or implementation shortfall is derived
below with the understanding that we need to use the Expectation operator
when we are working with estimates or future prices. (Kissell 2006)
provides more details including the formula where the portfolio may
be partly executed. The list of symbols we use are,
\end{doublespace}
\begin{itemize}
\begin{doublespace}
\item $\bar{S}$, the total number of shares that need to be traded.
\item $T$, the total duration of trading.
\item $N$, the number of trading intervals.
\item $\tau=T/N$, the length of each trading interval. We assume the time
intervals are of the same duration, but this can be relaxed quite
easily. In continuous time, this becomes, $N\rightarrow\infty,\tau\rightarrow0$.
\item The time then becomes divided into discrete intervals, $t_{k}=k\tau,\;k=0,...,N$.
\item For simplicity, let time be measured in unit intervals giving, $t=1,2,...,T$.
\item $S_{t}$, the number of shares acquired in period $t$ at price $P_{t}$.
\item $P_{0}$ can be any reference price or benchmark used to measure the
slippage. It is generally taken to be the arrival price or the price
at which the portfolio manager would like to complete the purchase
of the portfolio.
\item Any trading trajectory, would look to formulate an optimal list of
total pending shares, $W_{1},...,W_{T+1}$. Here, $W_{t}$ is the
number of units that we still need to trade at time $t$. This would
mean, $W_{1}=\bar{S}$ and $W_{T+1}=0$ implies that $\bar{S}$ must
be executed by period $T$. Clearly, $\bar{S}=\underset{j=1}{\overset{T}{\sum}}S_{j}$.
This can equivalently be represented by the list of executions completed,
$S_{1},...,S_{T}$. Here, $W_{t}=W_{t-1}-S_{t-1}$ or $S_{t-1}=W_{t-1}-W_{t}$
is the number of units traded between times $t-1$ and $t$. $W_{t}$
and $S_{t}$ are related as below.
\[
W_{t}=\bar{S}-\underset{j=1}{\overset{t-1}{\sum}}S_{j}=\underset{j=t}{\overset{T}{\sum}}S_{j}\qquad,t=1,...,T.
\]
Using the above notation, 
\end{doublespace}
\end{itemize}
\begin{doublespace}
\[
\text{Paper Return}=\bar{S}P_{T}-\bar{S}P_{0}
\]
\[
\text{Real Portfolio Return}=\bar{S}P_{T}-\left(\sum_{t=1}^{T}S_{t}P_{t}\right)
\]
\begin{eqnarray*}
\text{Implementation Shortfall} & = & \text{Paper Return}-\text{Real Portfolio Return}\\
 & = & \left(\sum_{t=1}^{T}S_{t}P_{t}\right)-\bar{S}P_{0}
\end{eqnarray*}
The innovation we introduce would incorporate our earlier discussion
about breaking the total impact or slippage, Implementation Shortfall,
into the part from the participants own decision process, Market Impact,
and the part from the decision process of all other participants,
Market Timing. This Market Impact, would capture the actions of the
participant, since at each stage the penalty a participant incurs
should only be the price jump caused by their own trade and that is
what any participant can hope to minimize. A subtle point is that
the Market Impact portion need only be added up when new price levels
are established. If the price moves down and moves back up (after
having gone up once earlier and having been already counted in the
Impact), we need not consider the later moves in the Market Impact
(and hence implicitly left out from the Market Timing as well). This
alternate measure would only account for the net move in the prices
but would not show the full extent of aggressiveness and the push
and pull between market participants and hence is not considered here,
though it can be useful to know and can be easily incorporated while
running simulations. Our measure of the Market Impact, for a buy order,
then becomes,
\[
\text{Market Impact}=\sum_{t=1}^{T}\left\{ \max\left[\left(P_{t}-P_{t-1}\right),0\right]S_{t}\right\} 
\]
The Market Timing is then given by,
\begin{eqnarray*}
\text{\text{Market Timing}} & = & \text{Implementation Shortfall}-\text{\text{Market Impact}}\\
 & = & \left(\sum_{t=1}^{T}S_{t}P_{t}\right)-\bar{S}P_{0}-\sum_{t=1}^{T}\left\{ \max\left[\left(P_{t}-P_{t-1}\right),0\right]S_{t}\right\} 
\end{eqnarray*}

\end{doublespace}
\begin{doublespace}

\subsection{Deep Dive Results}
\end{doublespace}

\begin{doublespace}
We perform many levels of comparisons and tests to gauge the impact
of the changes. Our information set for the deeper dive consists of
two datasets. One is the daily close price, average spread, total
volume and total number of trades across each of the 100 securities
starting from July-01-2013 to Dec-10-2014. The other dataset comprises
orders on these securities for the same time period. This information
set contains all the standard order level information like number
of shares, value, number of executions taken to fill the order and
also includes Market Impact, Timing and the Total Slippage. All the
variables can be measured based on observations done on an exchange,
since our study is structured as a self contained closed system, except
for the inclusion of the FX rate, which is required to construct notional
buckets in USD and helps relate to a broader audience and to facilitate
inferences to be drawn easily. All the results are depicted using
summary tables and graphical elements in section \ref{sec:Appendix---II},
Appendix - II.

\begin{figure}[H]
\includegraphics[width=18cm,height=18cm]{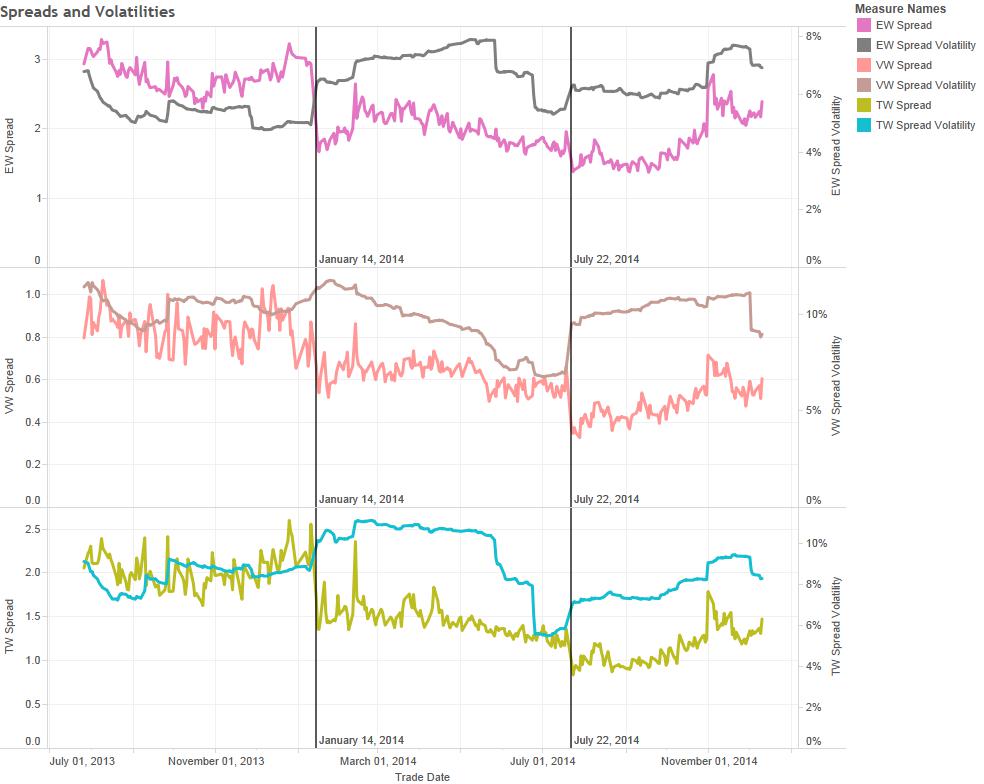}

\caption{Spreads and Volatilities}
\label{Spreads and Volatilities}
\end{figure}

\begin{figure}[H]
\includegraphics[width=18cm,height=20cm]{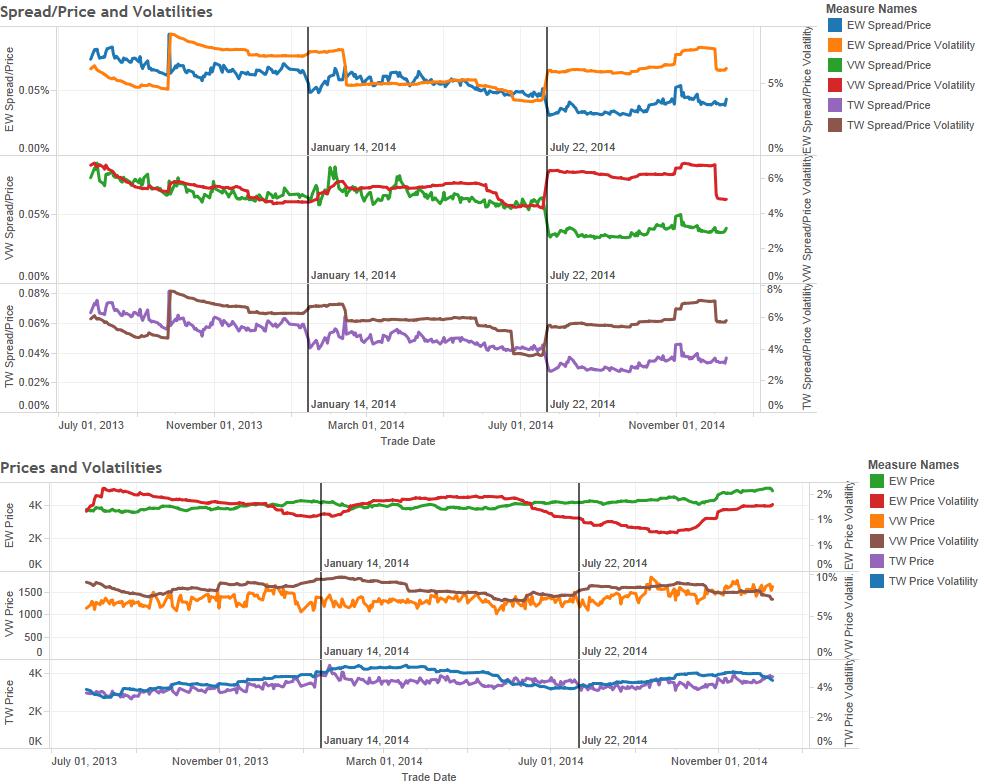}

\caption{Spread/Price, Price and Volatilities}
\label{Spread/Price and Volatilities}
\end{figure}

\begin{figure}[H]
\includegraphics[width=18cm,height=20cm]{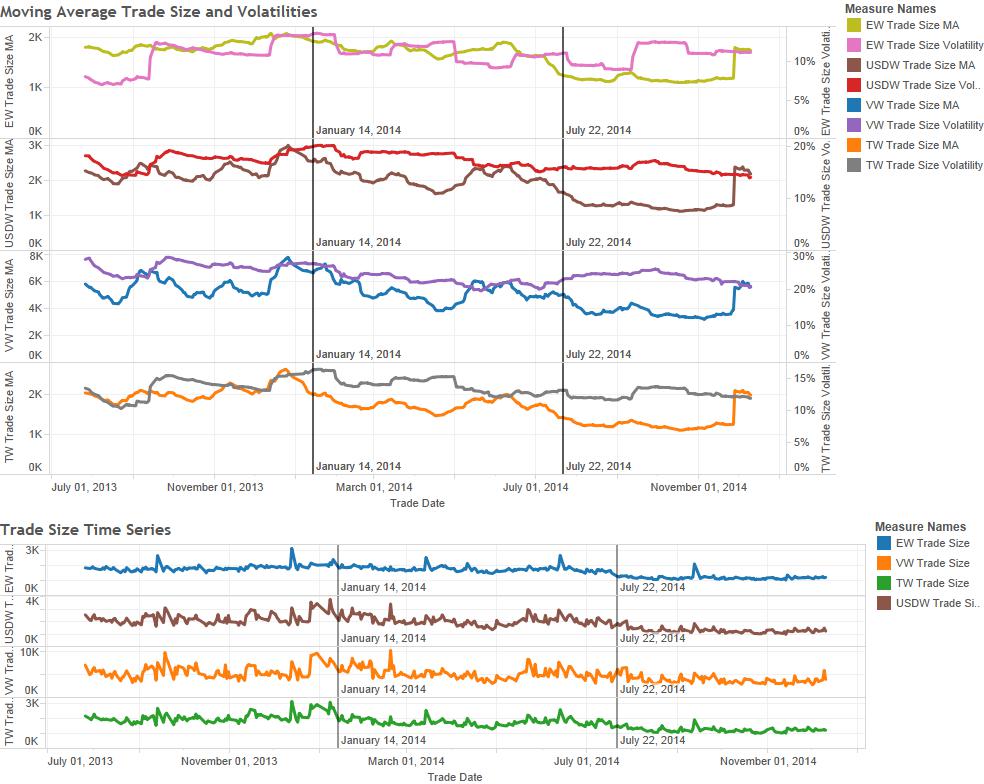}

\caption{Trade Size and Volatilities}
\label{Trade Size and Volatilities}
\end{figure}

\begin{figure}[H]
\includegraphics[width=18cm]{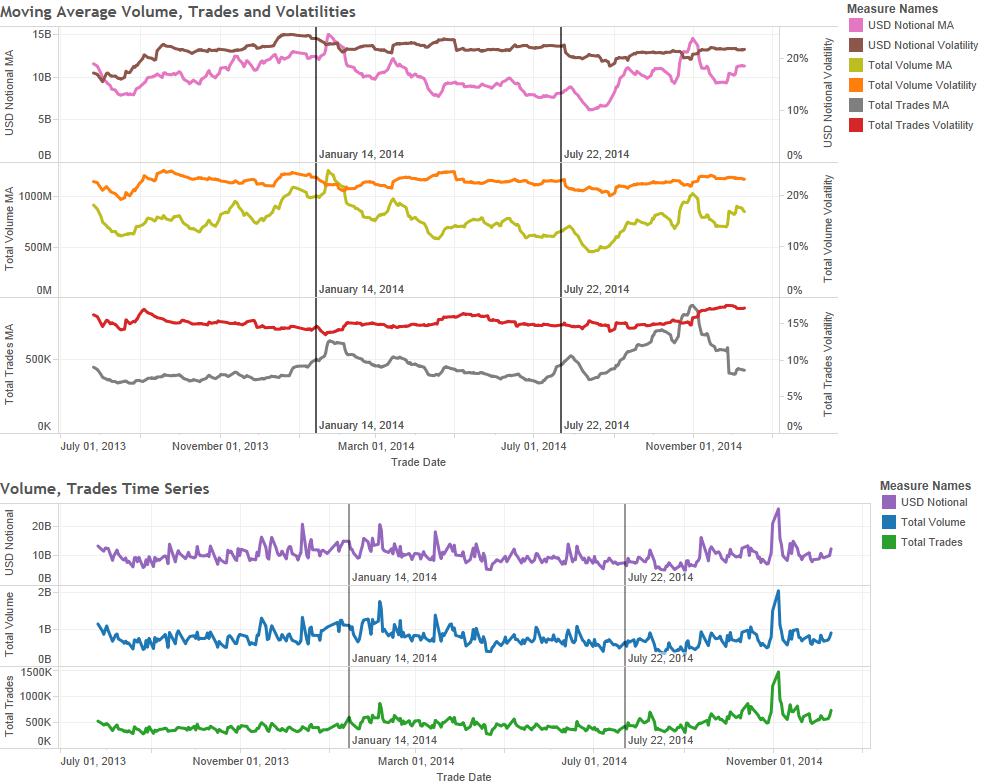}

\caption{Volume, Trades and Volatilities}
\label{Volume and Volatilities}
\end{figure}

We first calculate the Equal Weighted, Volume Weighted and Trade Weighted
Spread, Prices, the ratio of the Spread and Price and the Trade Size.
We use the average spread, the close price, the total daily volume
and the total number of daily trades at the security level. The fall
in the spread, both the average spread and the ratio of the spread
by the price, and the trade size around the two event days is easily
seen in the time series graph (Figures \ref{Spreads and Volatilities},
\ref{Spread/Price and Volatilities} and \ref{Trade Size and Volatilities}).
The trend in the trade size is better inferred when we smooth it using
a ten day moving average filter, being conscious of the fact there
will be a lag before we observe the values going down. We supplement
all the individual variables with the 90 day moving volatility of
each of the time series. The volatilities of the spread moves upward
around the two event days, but we cannot conclude that a new higher
volatility level is established. When we consider the price volatility
(Figure \ref{Spread/Price and Volatilities}), it is not clearly evident
that volatility has trended upwards. This is also not clearly established
from the volatility time series at the security level, hence we do
not report the security level volatilities. Both the raw values and
filtered values for the volumes and the number of trades, do not show
any discernible trend (Figure \ref{Volume and Volatilities}). A point
to bear in mind is that these events would have a greater effect on
intraday volatility and this is something to be checked for in later
studies. Greater volatility results in more efforts at managing a
more uncertain environment. In addition to affecting the Market Impact
numbers, this would be reflected in the Market Timing as well and
hence in the overall Slippage numbers.

\begin{figure}[H]
\includegraphics[width=18cm,height=22cm,keepaspectratio]{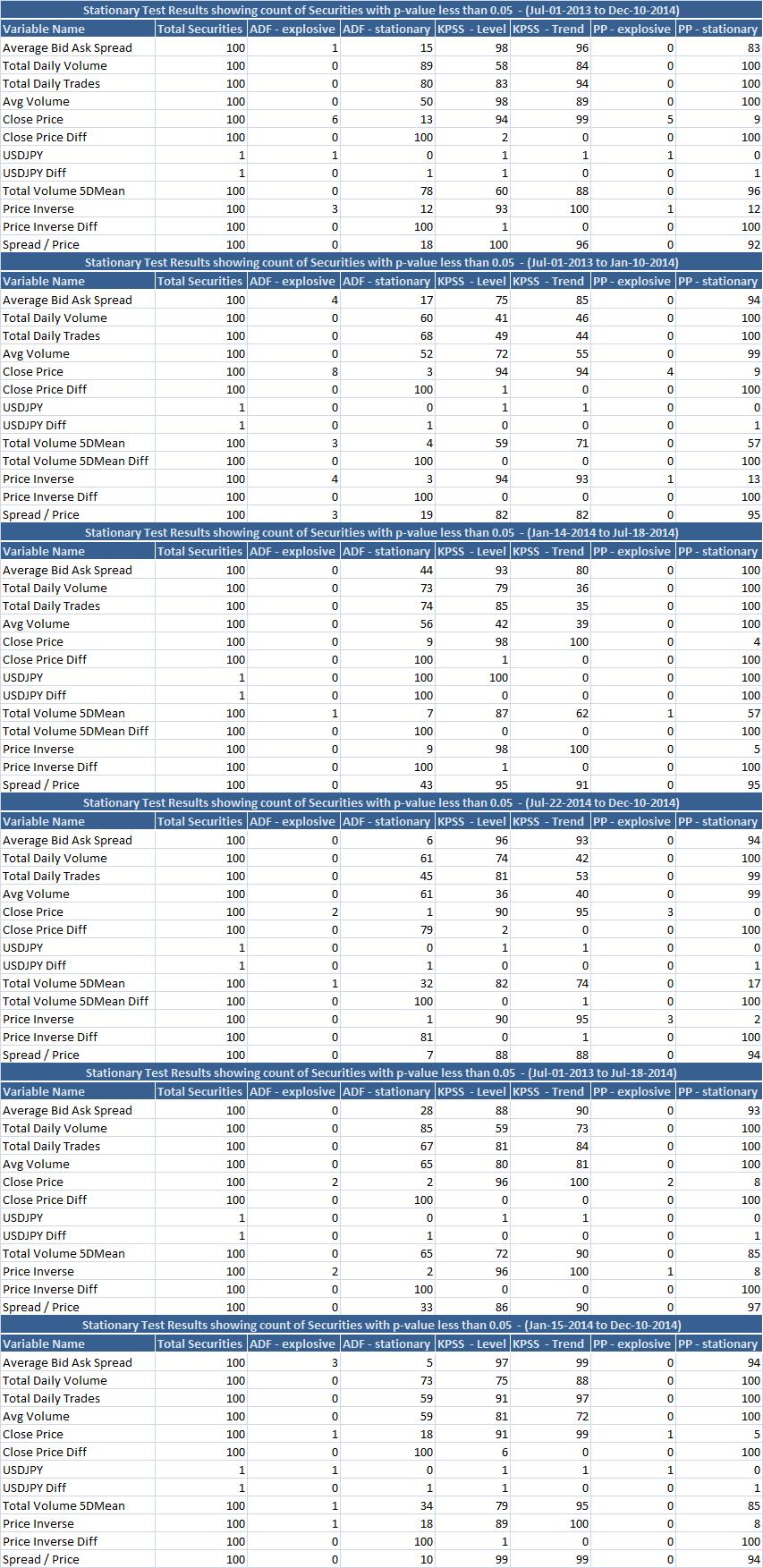}

\caption{Stationary Test Results}
\label{Stationary Test Results}
\end{figure}

\begin{figure}[H]
\includegraphics[width=9cm,height=20cm,keepaspectratio]{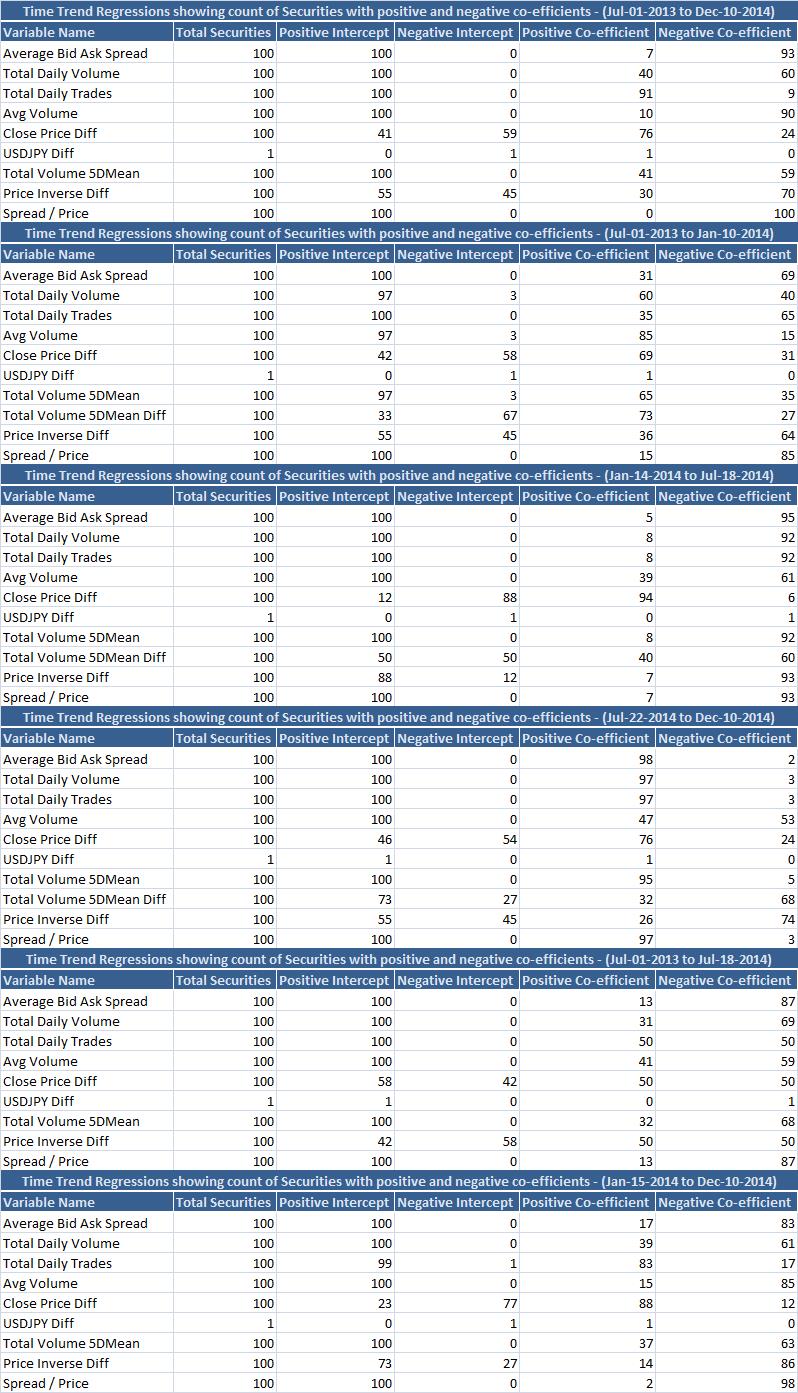}\includegraphics[width=9cm,height=20cm,keepaspectratio]{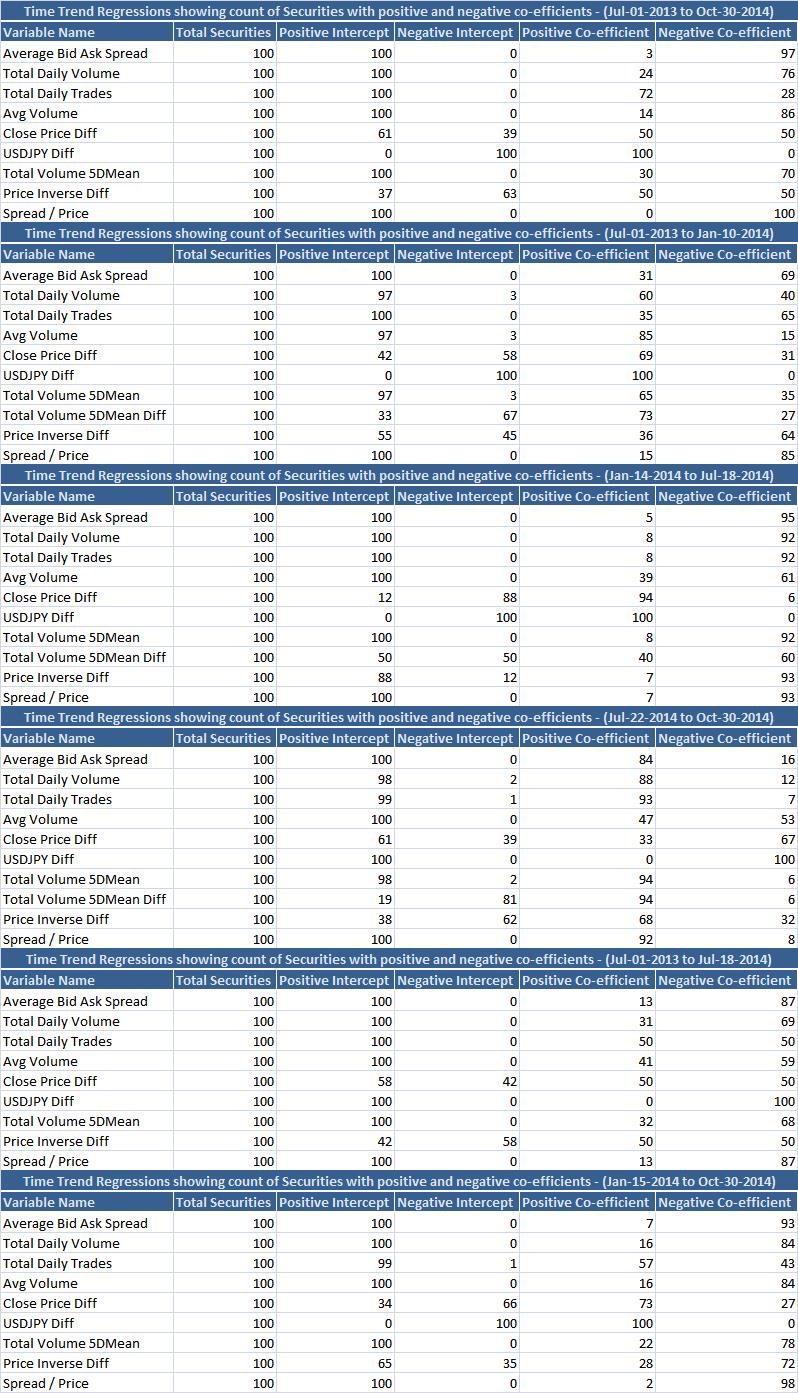}

\caption{Time Trend Regression Results}
\label{Time Trend Regression Results}
\end{figure}

We perform standard stationary tests on price, volume, spread and
other variables at the individual security level. We employ the Augmented
Dickey-Fuller (ADF) Test, the KPSS test and Phillips-Perron (PP) test.
The null hypothesis for the ADF and PP test is that there is a unit
root against the alternate that the series is explosive or stationary.
The KPSS null hypothesis is that the series is level or trend stationary
against the alternate that there is a unit root. It is easily apparent
that total daily volume, daily average spread, average volume, ratio
of spread by price and the number trades are stationary. Prices, Inverse
of the Price and USD/JPY FX rates are not. We repeat these tests across
the below six samples that we create from the overall dataset. We
see that the first difference of the non stationary variables results
in a stationary time series. We report the count of securities with
a p-value less than 0.05 in Figure \ref{Stationary Test Results}.
In all our regressions, we include the first difference of the variables
which are non stationary.
\end{doublespace}
\begin{enumerate}
\begin{doublespace}
\item Sample Full , SF: The entire dataset, from Jul-01-2013 to Dec-10-2014.
\item Sample One, S1: The start of the dataset to the first event, from
Jul-01-2013 to Jan-10-2014.
\item Sample Two, S2: The first event to the second event, from Jan-14-2014
to Jul-18-2014.
\item Sample Three, S3: The second event to the end of the dataset, from
Jul-22-2014 to Dec-10-2014.
\item Sample Four, S4: From the start of the dataset to the second event,
from Jul-01-2013 to Jul-18-2014.
\item Sample Five, S5: From the first event to the end of dataset, from
Jan-15-2014 to Dec-10-2014.
\end{doublespace}
\end{enumerate}
\begin{doublespace}
Next, we fit a trend line with a non-zero intercept across each of
the variables at the security level and count the number of securities
that show an increasing trend (Figure \ref{Time Trend Regression Results}
summarizes the results for each of the variables, $z_{t}$). This
is also equivalent to checking a deterministic time trend in the variables
as shown in (Eq: \ref{eq:Time-Trend}). 
\begin{equation}
z_{t}=\beta_{0}+\beta_{1}t+\varepsilon_{t}\label{eq:Time-Trend}
\end{equation}
It is clearly seen from this that spreads have come down and the average
volume has decreased consistently across most of the names around
the two events. To clarify an apparent divergence, the counts for
the third sample show a high number for increased spread, but this
sample includes only days after the second event and the fall in spreads
have occurred before the start of this sample. There is a jump in
the spread around Oct-31-2014 (also the volume, number of trades and
notational) which causes the spread trend lines to have increased
slope in the last sample period, overcoming the effect of the earlier
decrease on Jul-22-2014. Barring this outlier, which was caused by
a sudden surge in prices, possibly attributable to the Bank of Japan
unexpectedly adding stimulus by targeting a \$726 billion USD annual
expansion in the central bank's monetary base and the \$1.2 trillion
USD Government Pension Investment Fund announcing plans to more than
double its target allocation to Japanese stocks to 25 percent of assets,
(See End-notes \ref{enu:Japanese-stocks-soared,} and \ref{enu:Japanese-shares-swing})
the results are consistent and as expected. To supplement the above
six samples, whenever the sample includes the last date Dec-10-2014,
we run an extra set of regressions until Oct-30-2014, allowing us
to judge the results after removing the effects of this abnormal jump.

\begin{figure}[H]
\includegraphics{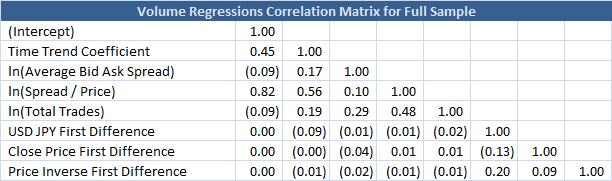}

\caption{Volume Regression Correlation Matrix}
\label{Volume Regression Correlation Matrix}
\end{figure}

\begin{figure}[H]
\includegraphics[width=18cm]{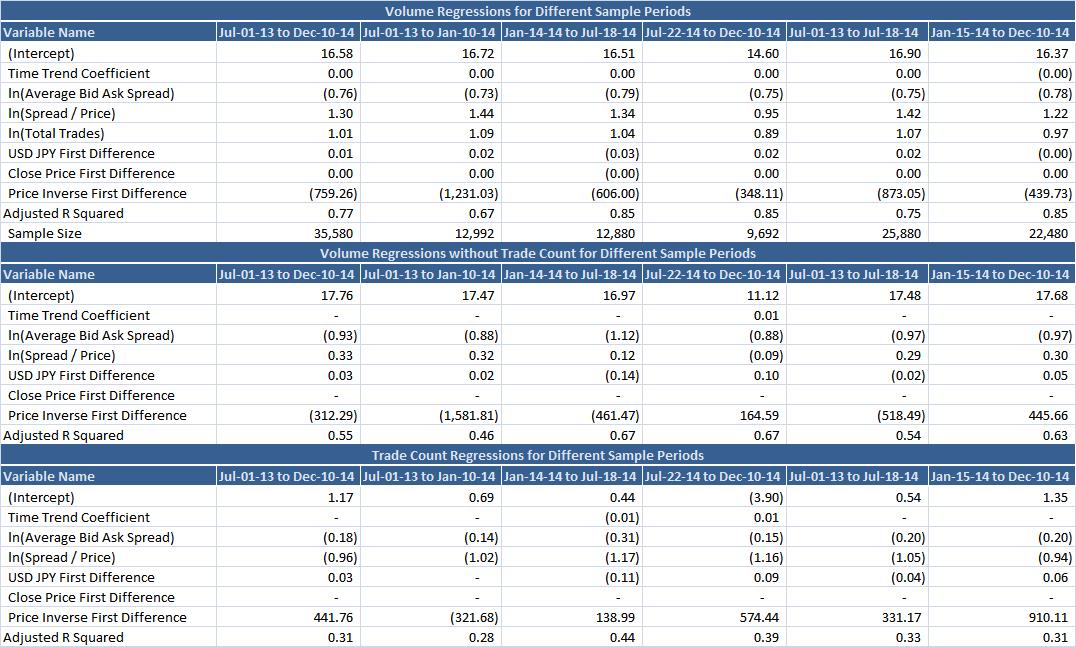}

\caption{Volume Regression Results}
\label{Volume Regressions}
\end{figure}

The volume and number of trades trend is inconclusive just by looking
at the slope of the time trend. Hence to assess this further, we run
some regressions across each of the six samples. We run three sets
of regressions. In the first (Eq: \ref{eq:Volume-Regression-One};
Figure \ref{Volume Regressions} shows all the variables and summarizes
the results), the volume is the dependent variable, $y_{t}$. Spread
and Number of Trades are the key independent variables. In the second,
we exclude the number of trades. The third regression is similar to
the second except that we set the number of trades as the dependent
variable. 
\begin{align}
y_{t} & =\beta_{0}+\beta_{1}t+\beta_{2}\ln\left(AvgBidAskSpread_{t}\right)+\beta_{3}\ln\left(\frac{Spread_{t}}{Price_{t}}\right)+\beta_{4}\ln\left(TotalTrades_{t}\right)\label{eq:Volume-Regression-One}\\
 & +\beta_{5}\ln\left(USDJPYFirstDifference_{t}\right)+\beta_{6}\ln\left(ClosePriceFirstDiff_{t}\right)+\beta_{7}\ln\left(\frac{1}{ClosePriceFirstDiff_{t}}\right)+\varepsilon_{t}
\end{align}
Here, the ratio of the spread by price and the inverse of the price
act as control variables. We also include the USD/JPY as an additional
control variable. All the variables except the FX rate are significant.
The correlation matrix is in Figure \ref{Volume Regression Correlation Matrix}.
It is clearly seen from the regression coefficients (Figure \ref{Volume Regressions})
that the volume and number of trades increase when spreads fall. We
get similar results when we take the lag of the independent variables
by one day and by one week.

\begin{figure}[H]
\includegraphics{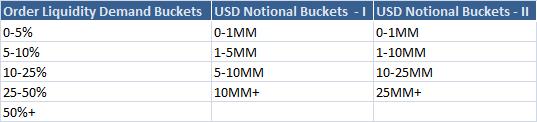}

\caption{Order Liquidity Demand and USD Notional Size Buckets}
\label{Order Buckets}
\end{figure}

\begin{figure}[H]
\includegraphics[width=18cm,height=18cm]{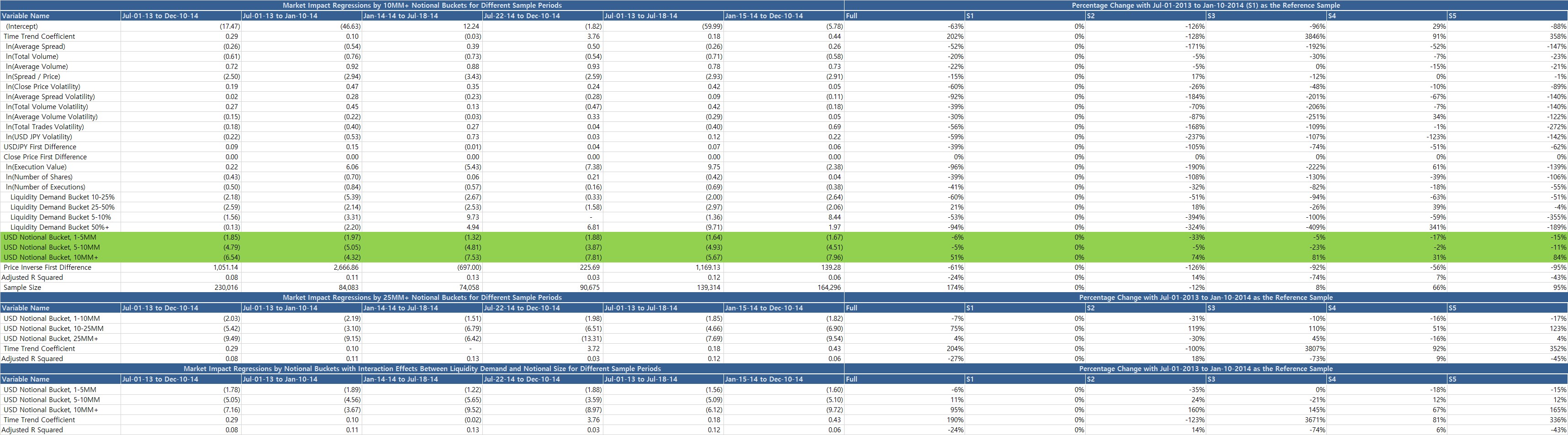}

\caption{Trading Cost Regression Results}
\label{Cost Regressions}
\end{figure}

\begin{figure}[H]
\includegraphics[width=17cm]{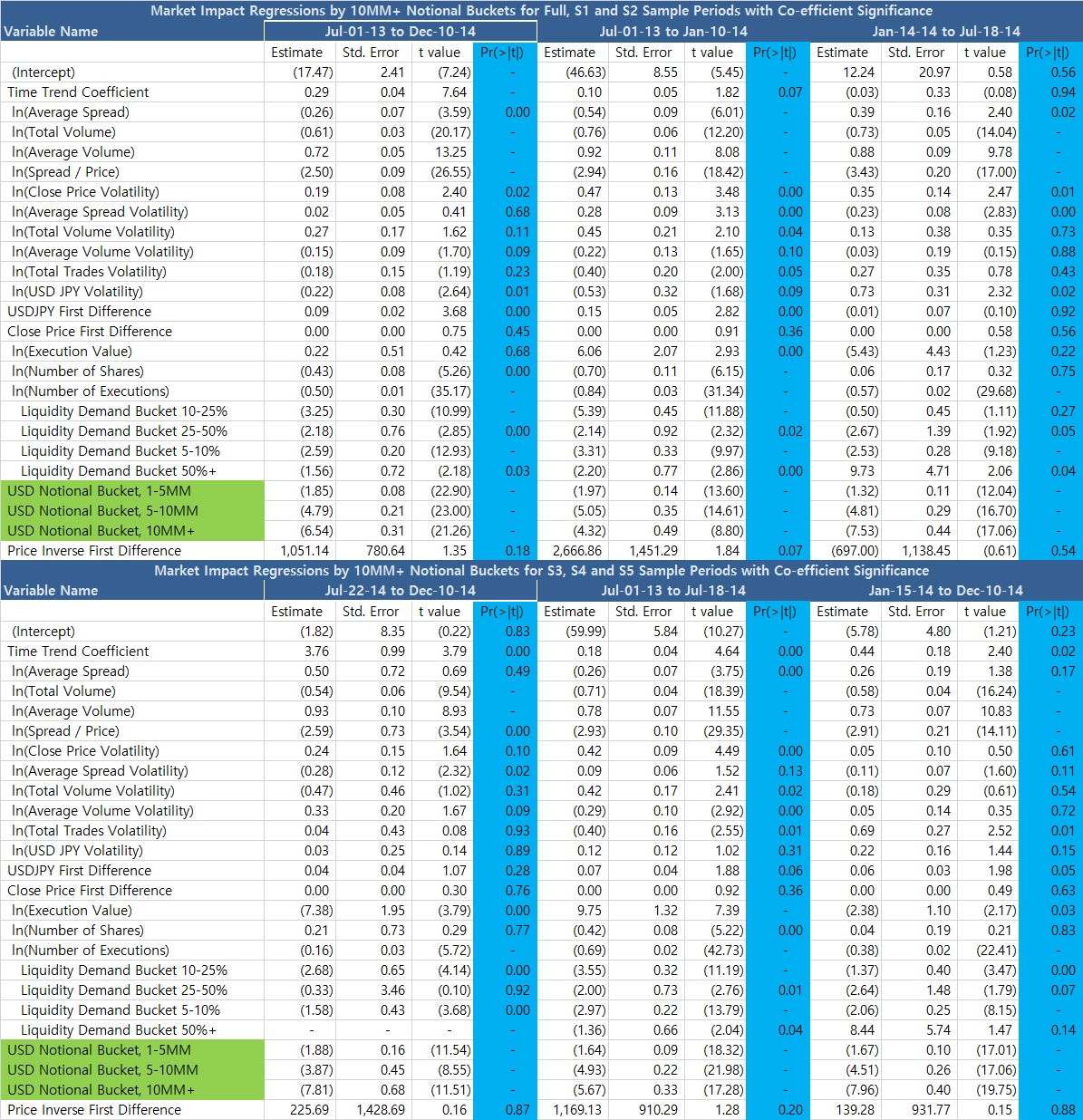}

\caption{\label{fig:Cost-Regression-Coefficient}Cost Regression Coefficient
Significance}

\end{figure}

Saving the best for last, we look at trading costs. We run separate
regressions with all three of our cost metrics described earlier,
Market Impact, Market Timing and Total Slippage as the dependent variables.
We find that, the results are similar across all three proxies of
the trading cost, but the adjusted R-Squared is higher with the Market
Impact, $MI_{t}$ (Eq: \ref{eq:MI-Cost-Regresson}; Figure \ref{Cost Regressions}
shows all the variables and summarizes the results; Figure \ref{fig:Cost-Regression-Coefficient}
gives the significance of the regression cofficients for different
sample periods). The real cost associated with a trader's effort in
seeking liquidity is given by the Market Impact, hence we report and
discuss only those results. We need to interpret the results keeping
in mind that trading costs are notoriously difficult to predict, and
models relating costs to other variables come with a high level of
variance. We wish to understand how the trading costs have changed
across orders of different sizes given by the independent variable,
the USD notional, bucketed into four categories. We repeat the regressions
for two different sets of categorizations of the Notional buckets
(Figure \ref{Order Buckets}). With such a setup, the orders in the
smallest notional bucket, 0-1MM (million) USD, become the benchmark
against which we measure the trading costs in the other buckets. 
\end{doublespace}

\begin{figure}[H]
\includegraphics[width=18cm,height=14cm]{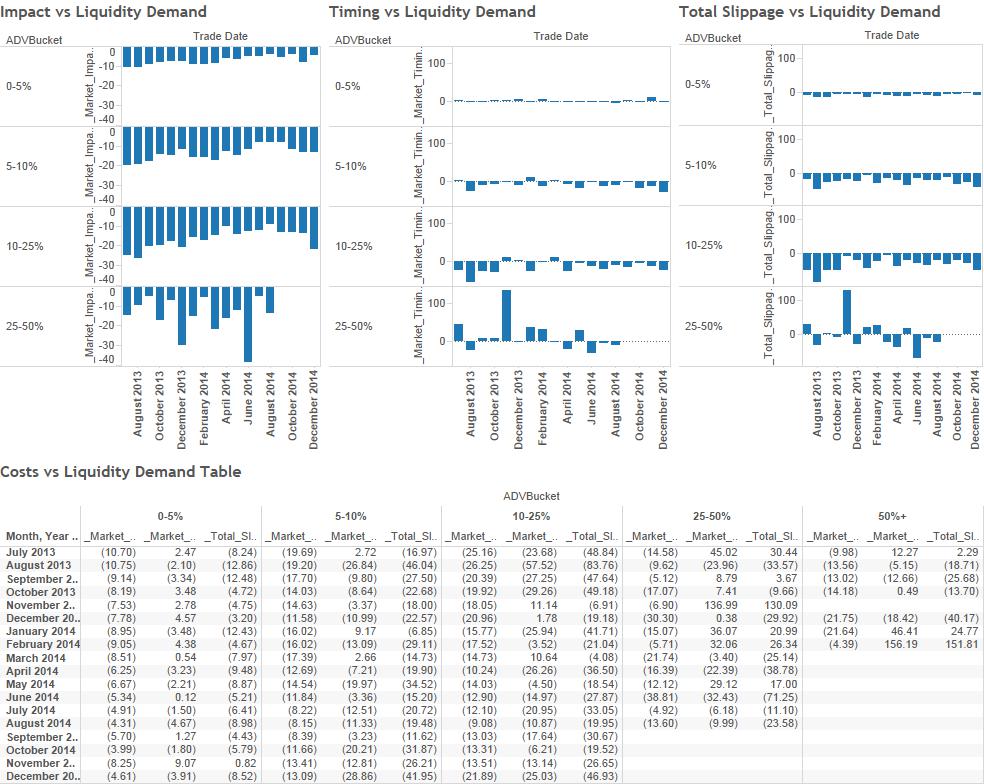}

\includegraphics[width=18cm,height=9.8cm]{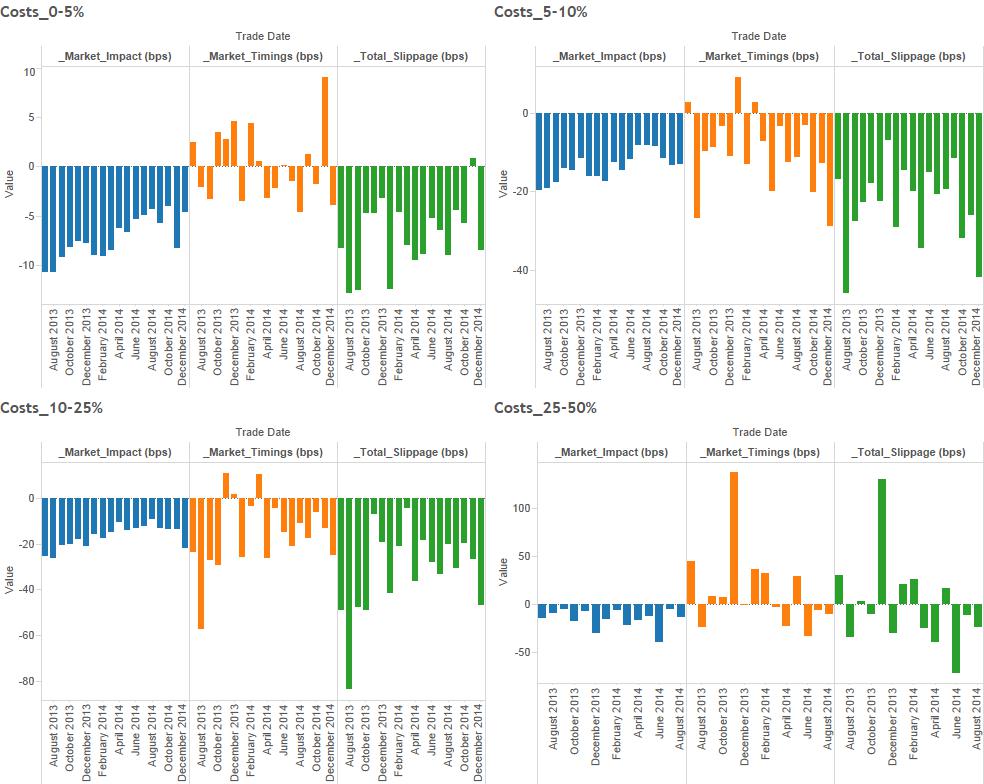}

\caption{Trading Costs by Liquidity Demand}
\label{Costs By Liquidity Demand}
\end{figure}

\begin{figure}[H]
\includegraphics[width=18cm,height=14cm]{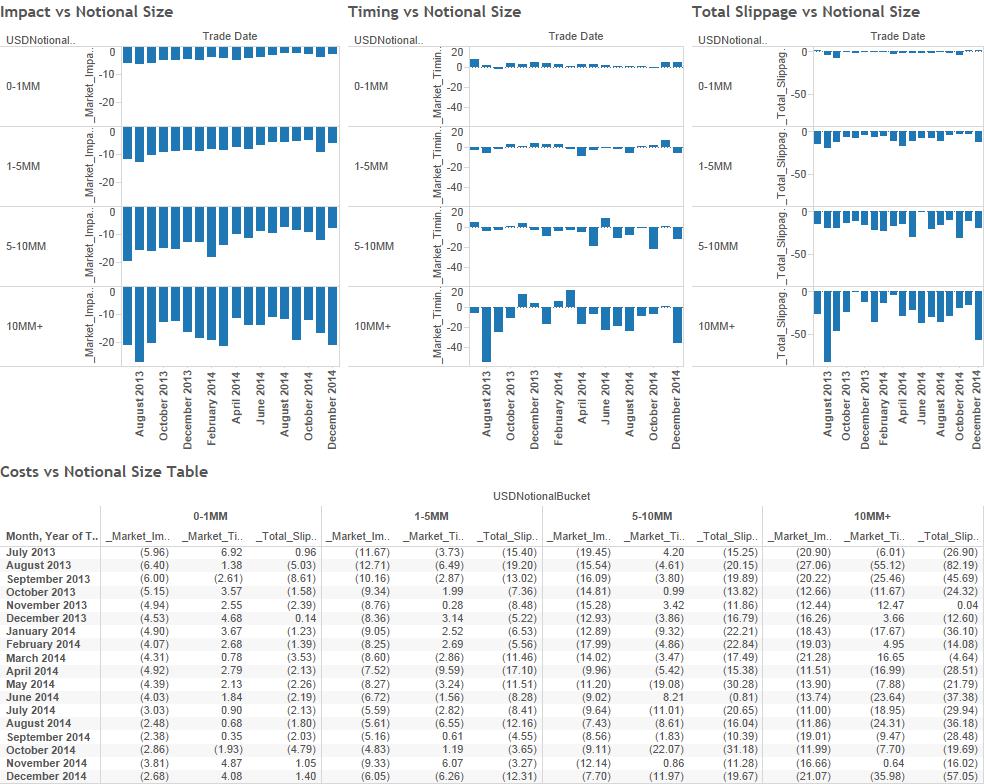}

\includegraphics[width=18cm,height=9.8cm]{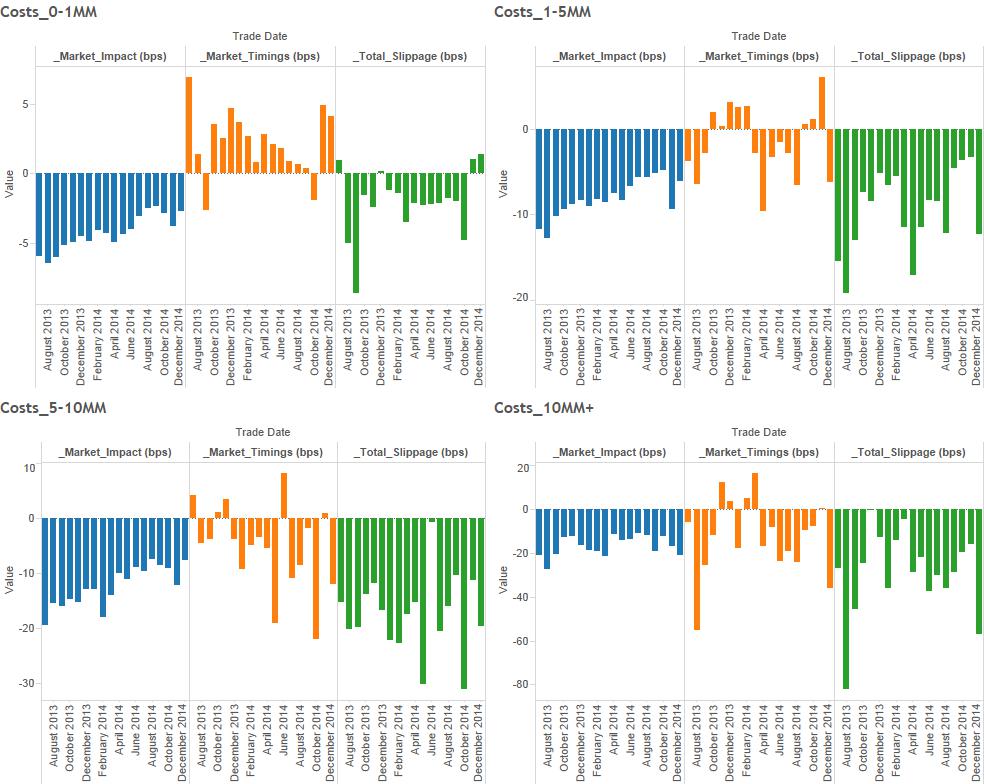}

\caption{Trading Costs by Notional Size - 10MM+}
\label{Costs by Notional 10MM+}
\end{figure}

\begin{figure}[H]
\includegraphics[width=18cm,height=14cm]{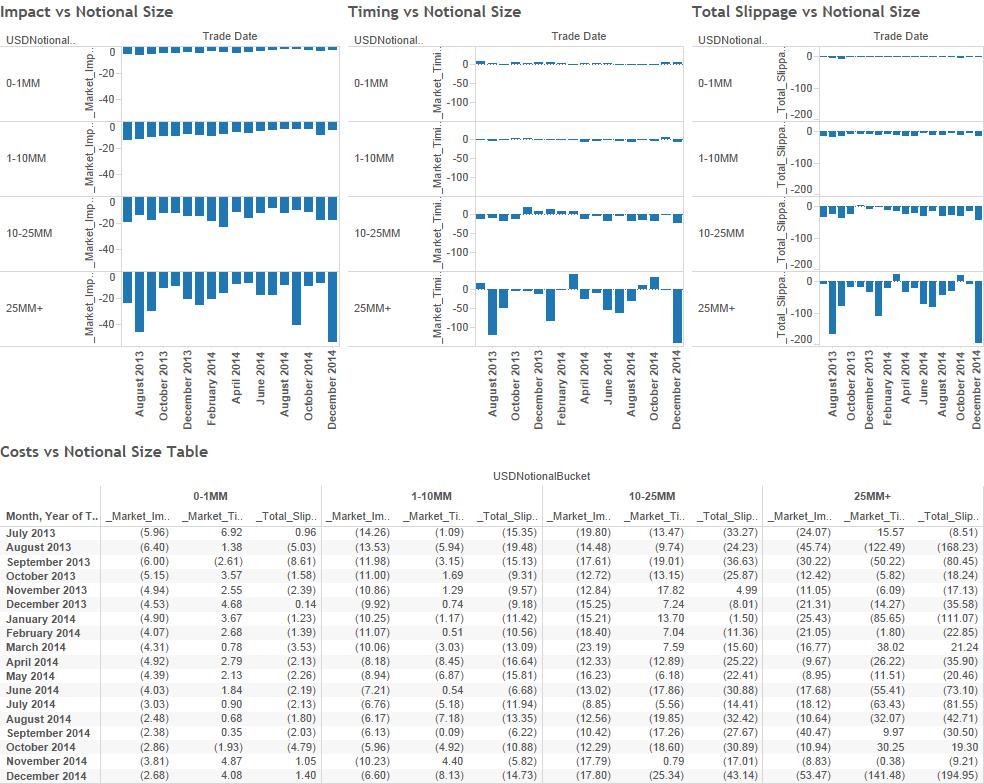}

\includegraphics[width=18cm,height=9.8cm]{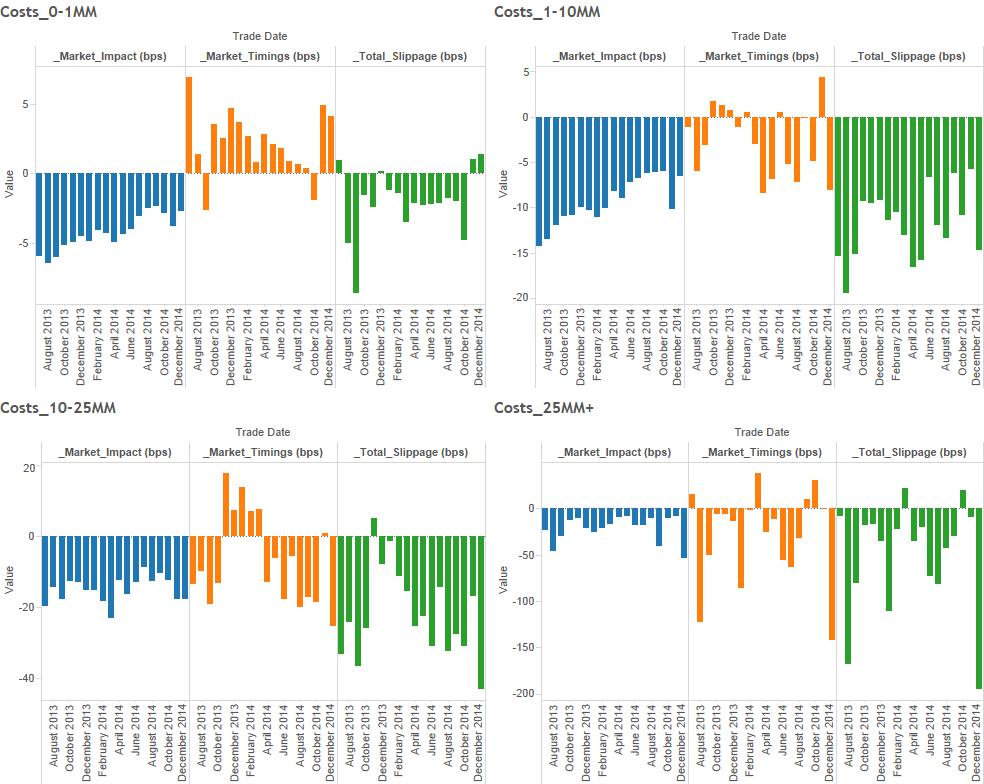}

\caption{Trading Costs by Notional Size - 25MM+}
\label{Costs by Notional 25MM+}
\end{figure}

\begin{figure}[H]
\includegraphics[width=18cm]{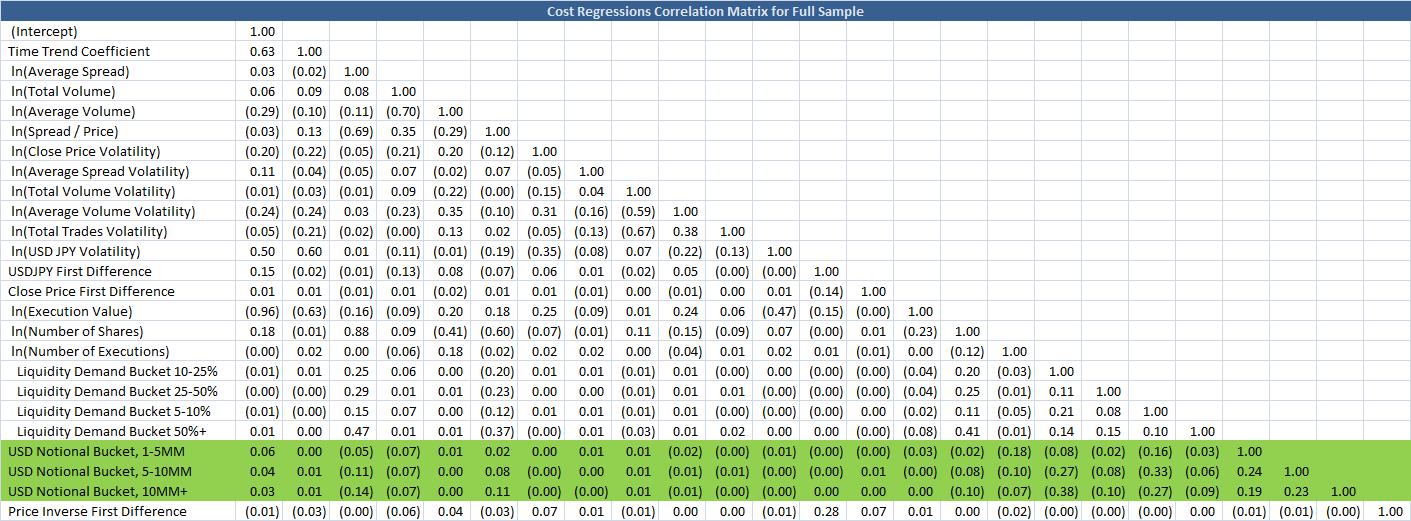}

\caption{Cost Regression Correlation Matrix}
\label{Cost Regression Correlation Matrix}
\end{figure}

We include a whole smattering of independent variables that act as
controls, including usual suspects such as spread, spread divided
by price, total trading volume, average trading volume, closing price
first difference, inverse of closing price first difference, number
of shares, first difference of USDJPY FX Rates, total executed value,
number of executions, moving 90 day volatilities of price, spread,
volume, number of trades and the FX rate. The correlation matrix is
in Figure \ref{Cost Regression Correlation Matrix}. As an additional
control variable, we include the liquidity demanded by the order as
a percentage of the total daily volume bucketed into five categories.
The costs are on a decreasing trend from the beginning of our sample
(Figures \ref{Costs By Liquidity Demand}, \ref{Costs by Notional 10MM+}
and \ref{Costs by Notional 25MM+}). Hence, later studies should try
to include explanatory variables to account for this phenomenon. We
see similar results when we repeat the regressions with many different
variations in terms of which  explanatory variables we include, the
use of the liquidity demand buckets, interaction effects betwen liquidity
demand and notional buckets and the use of logarithms to scale the
explanatory variables. We do not run these regression with a time
lag since we are primarily interested in the contemporaneous relationship
between changes in the variables, but lag effects are not to be ruled
out and can be pursued later.
\begin{align}
MI_{t} & =\beta_{0}+\beta_{1}t+\beta_{2}\ln\left(AvgBidAskSpread_{t}\right)+\beta_{3}\ln\left(\frac{Spread_{t}}{Price_{t}}\right)+\beta_{4}\ln\left(TotalVolume_{t}\right)\label{eq:MI-Cost-Regresson}\\
 & +\beta_{5}\ln\left(90DayMovingClosePriceVolatility_{t}\right)+\beta_{6}\ln\left(TotalVolumeVolatility_{t}\right)+...+\varepsilon_{t}
\end{align}

\begin{doublespace}
As primary evidence of increased trading costs, we see from the regression
results (Figure \ref{Cost Regressions}) that the costs for the 10MM+
notional bucket after the event are higher relative to the other buckets.
We ascertain this by calculating the difference between the coefficients
of the notional buckets for the sample periods after either of the
two changes have happened with the corresponding notional bucket coefficient
for the sample before either of the events have happened. For example,
when considering the difference in the coefficients of the S2, S3
and S5 samples with the S1 sample coefficient: the 10MM+ bucket coefficients
are higher by 74\%, 81\% and 84\% respectively; the 5-10MM bucket
coefficients are lower by 5\%, 23\% and 11\%. This effect is most
clearly seen when comparing the coefficients of S2, S5 with S1 across
increasing bucket sizes. We can observe this effect in the graph of
the Market Impact by notional buckets over time (Figure \ref{Costs by Notional 10MM+}).
\end{doublespace}

\begin{figure}[H]
\includegraphics[width=17cm]{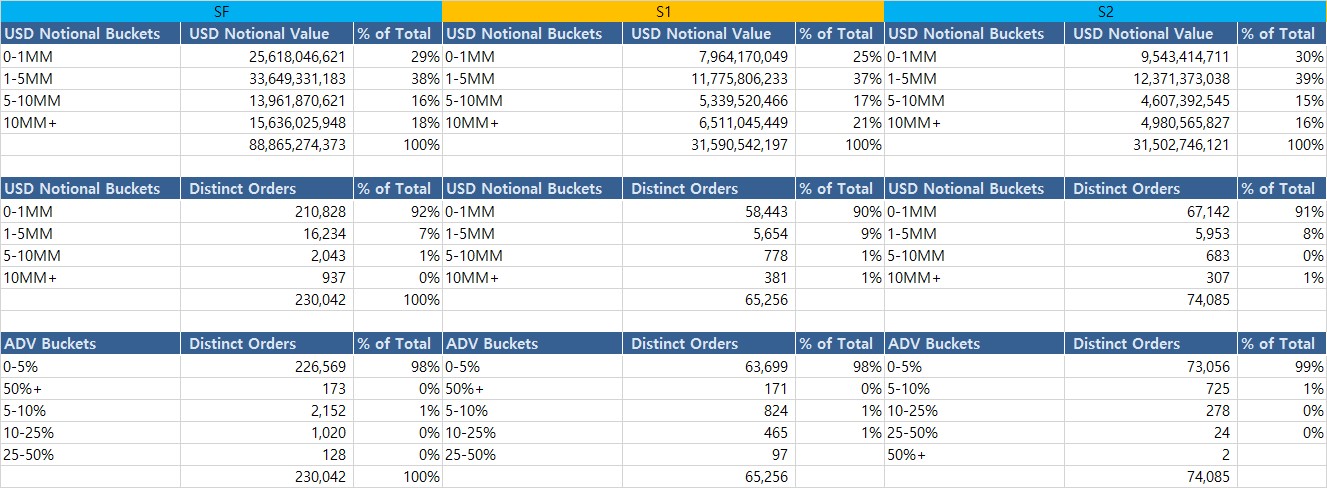}

\includegraphics[width=17cm]{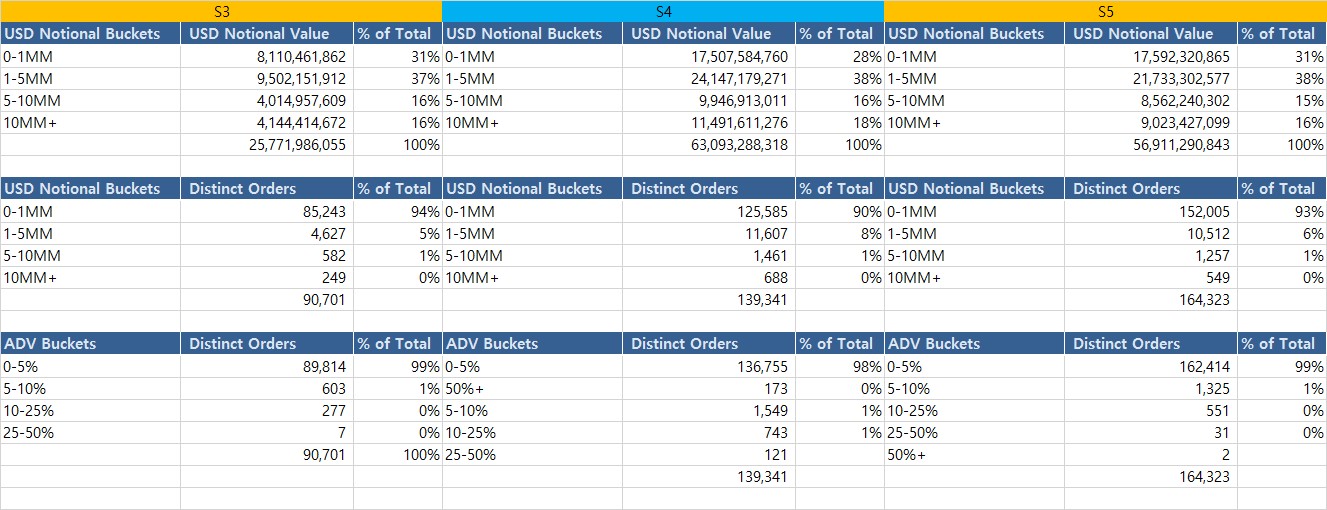}

\caption{\label{fig:USD-Notional-Buckets-10MM+}Order USD Notional Buckets
10MM+, Number of Orders and \% of Total Wealth}

\end{figure}

\begin{figure}[H]
\includegraphics[width=17cm]{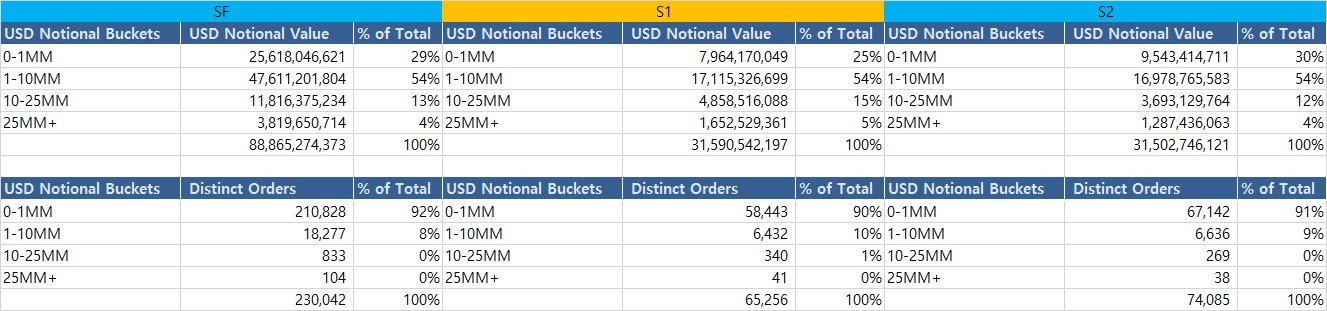}

\includegraphics[width=17cm]{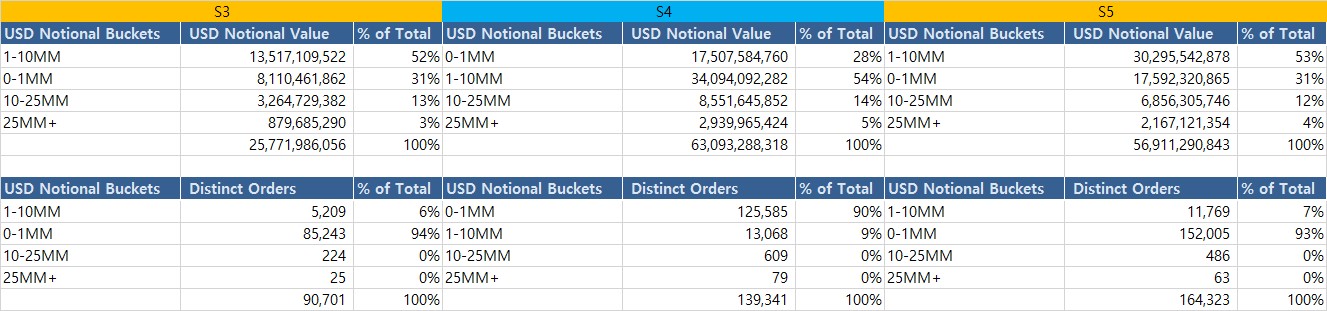}

\caption{\label{fig:USD-Notional-Buckets-25MM+}Order USD Notional Buckets
25MM+, Number of Orders and \% of Total Wealth}

\end{figure}

A justification for this choice of notional buckets is that such a
classification ensures we have a good number of orders and a good
percentage of the total notional value in each bucket (Figures \ref{fig:USD-Notional-Buckets-10MM+},
\ref{fig:USD-Notional-Buckets-25MM+} show the number of orders, the
notional USD and the percentage of totals for liquidity buckets and
notional buckets). Also, this classification is more intuitive to
grasp since it follows industry practice, especially used on many
trading desks, and helps to categorize order sizes as small, medium,
large and very large orders. The sample size is much smaller for the
the 25MM+ bucket than the 10MM+ buckets and hence the results are
not as reliable, but we include it for completeness. In the other
set of regressions with 25MM+ notional size categorization, for the
S2, S3 and S5 samples, the 10-25MM bucket coefficients are higher
by 119\%, 110\% and 123\% as compared to the 1-10MM buckets coefficients
which are lower by 31\%, 10\% and 17\% respectively. Including interaction
effects between the liquidity demand and notional buckets or excluding
the liquidity demand buckets does not improve or change the results
significantly.

In (Appendix \ref{sec:Appendix---III}) we provide a mathematical
justification for why endogeneity (Hamilton 1994; Gujarati 1995; End-note
\ref{enu:Endogeneity-is-a}) will not cause a major bias in our interpretation
of the coefficients of the notional buckets in the cost regressions
(Eq: \ref{eq:MI-Cost-Regresson}). Endogeneity can set in under three
scenarios: 1) if the dependent variables can influence the explanatory
variables and vice versa; 2) some key variables are omitted in the
regression model; 3) and there are errors in the measurement of the
variables. 

In our case, inconsistent estimates of the notional bucket coefficients
are a potential concern since market impact can influence price and
price can influence market impact. Clearly, in almost any study, it
is very hard to completely rule out omitted variables and measurement
errors. Specific to our study which is designed as a closed system,
as discussed earlier, there could be many potential candidate variables
that have been left out and any study of trading costs will have measurement
errors due to the high level of noise in the environment. Our justification
for why endogeneity causes no major issues in our setup is based on
the simple argument that we are looking at the changes in the coefficients
of the notional buckets before and after the event. Hence, under a
broad set of conditions, any bias in the coefficient estimates will
cancel out leaving us with the actual change in the coefficients giving
us a clear indication of how the costs have changed before and after
the event.

We wish to emphasize the rationale for structuring the regressions
to look at the change in the coefficients for order bucket size categories
versus looking at the regression co-efficient on a single indicator
of size (such as number of shares or order notional) and using a dummy
variable to capture orders before and after the event. This alternate
approach using dummy variables, tells us how the costs have changed
on the average order size before and after the event. But since any
trading data sample will be heavily skewed towards smaller orders
(Figures \ref{fig:USD-Notional-Buckets-10MM+}, \ref{fig:USD-Notional-Buckets-25MM+}),
we will end up measuring how costs have changed across smaller orders.
Since our study is trying to understand how costs have changed across
small and large orders, we would clearly need to have some categorization
of order sizes into buckets and a comparison of the corresponding
coefficients. 

\begin{figure}[H]
\includegraphics[width=17cm]{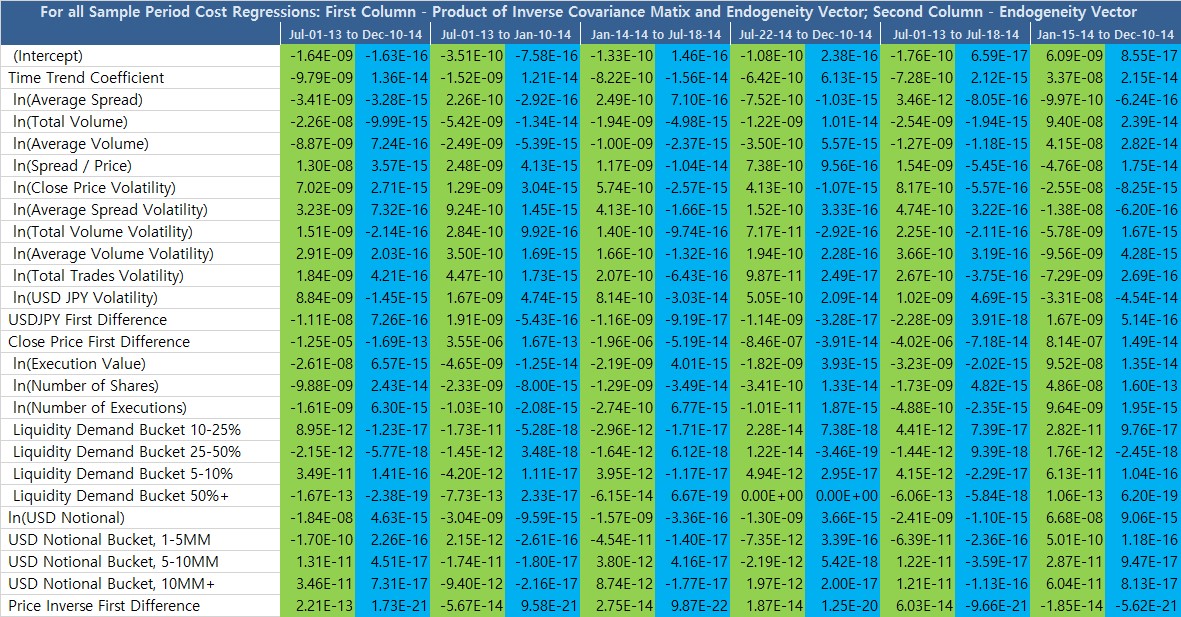}

\caption{\label{fig:Cost-Regression-Endogeneity}Cost Regression Endogeneity
Resolution}

\end{figure}

\begin{figure}[H]
\includegraphics[width=17cm]{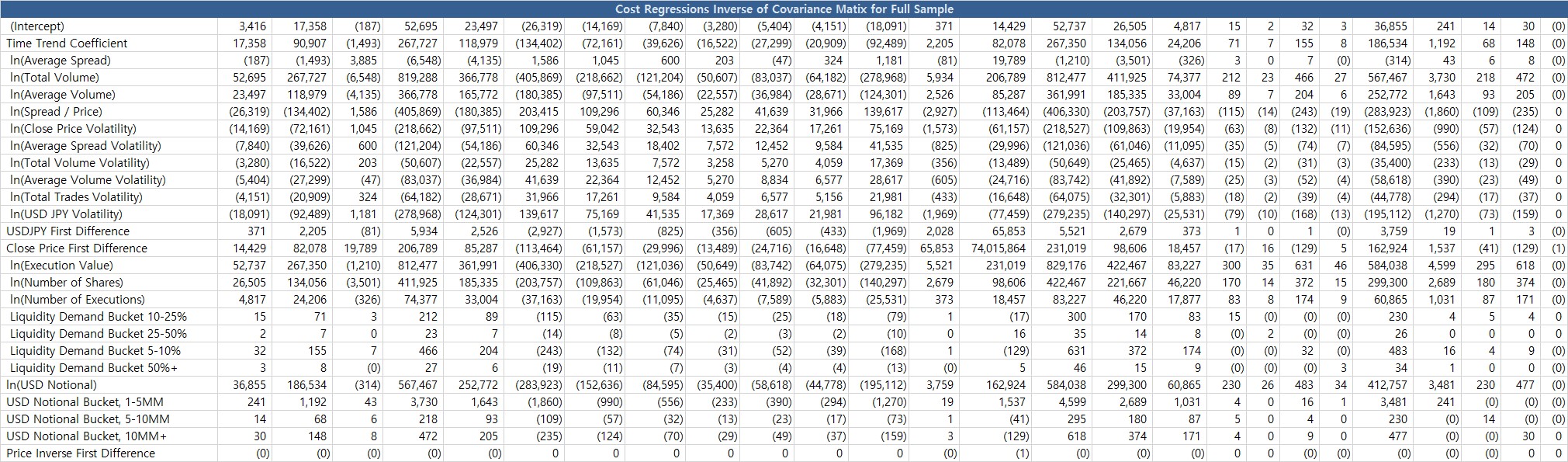}

\caption{\label{fig:Cost-Regression-Covariance}Cost Regression Covariance
Matrix Inverse}

\end{figure}

In addition, we are also including the closing price first difference,
inverse of the closing price first difference, the logarithm of the
total number of shares comprising the order and the logarithm of the
total executed value explicitly as explanatory variables. These adjustments
serve to minimize the influence of market impact on the price level
used to create the notional buckets. An additional adjustment, which
we do not employ, is to calculate the notional buckets for all the
orders based on the closing price and not the prices at which the
executions that comprise the order are transacted. Even without this
adjustment, we are able to rule out any bias in the coefficients as
illustrated in Figures \ref{fig:Cost-Regression-Endogeneity}, \ref{fig:Cost-Regression-Covariance}.
In Figure \ref{fig:Cost-Regression-Endogeneity} we show that the
product of the inverse of the co-variance matrix for the cost regressions
with the co-variance between the explanatory variables and the cost
regression residuals is very close to zero for all sample periods.
In Figure \ref{fig:Cost-Regression-Endogeneity} in the first column
we show the product of the inverse of the co-variance matrix for the
cost regressions with the co-variance between the explanatory variables
and the cost regression residuals. In the second column of Figure
\ref{fig:Cost-Regression-Endogeneity} we show the co-variance between
the explanatory variables and the cost regression residuals. In Figure
\ref{fig:Cost-Regression-Covariance} we show the inverse of the co-variance
matrix of the cost regression for the full sample period.

\begin{figure}[H]
\includegraphics[width=18cm,height=22cm]{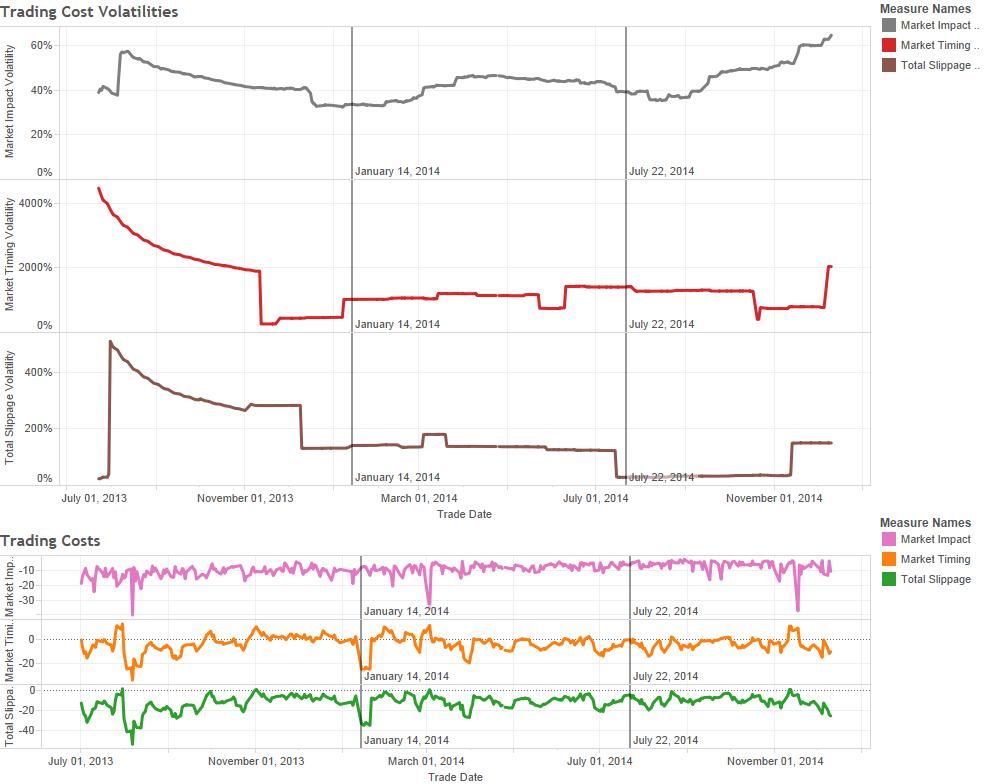}

\caption{Trading Costs and Volatilities}
\label{Trading Costs and Volatilities}
\end{figure}

\begin{figure}[H]
\includegraphics[width=18cm,height=10cm]{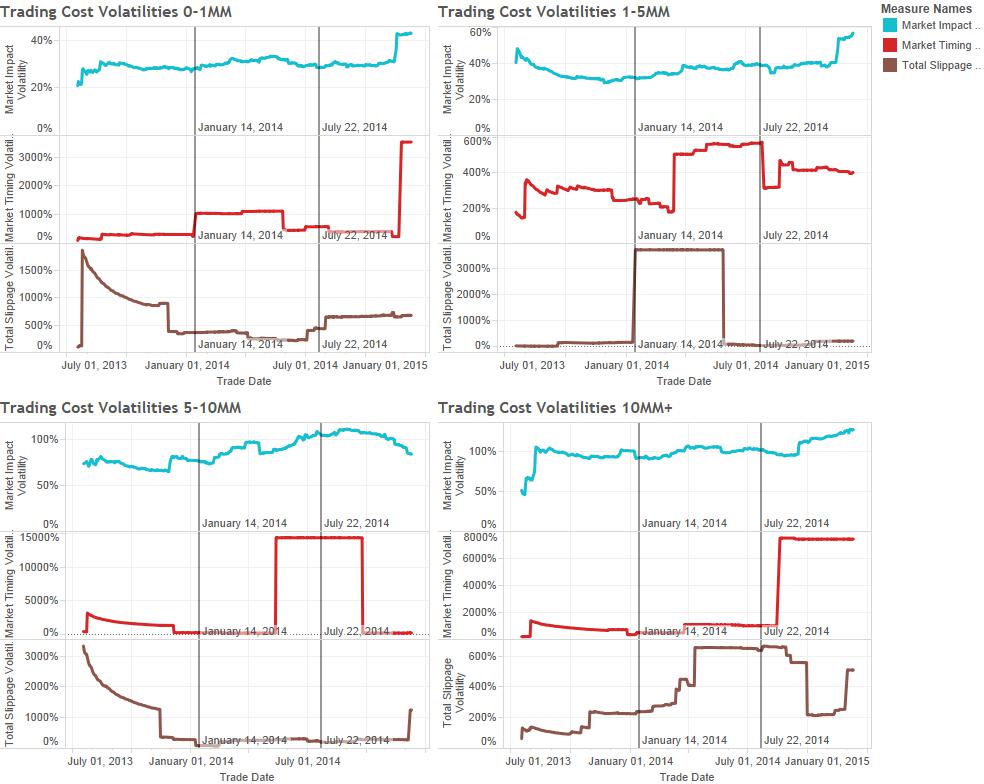}

\caption{Trading Cost Volatilities By Notional Buckets - 10MM+}
\label{Trading Cost Volatilities By Notional Buckets - 10MM+}
\end{figure}

\begin{figure}[H]
\includegraphics[width=18cm,height=10cm]{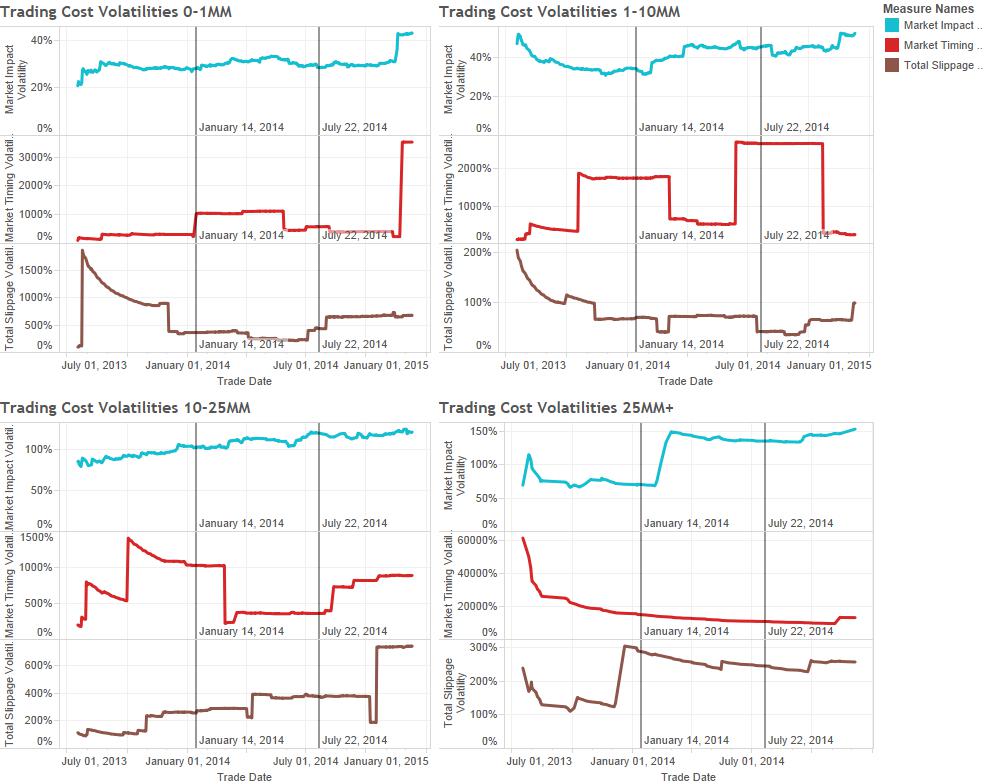}

\caption{Trading Cost Volatilities By Notional Buckets - 25MM+}
\label{Trading Cost Volatilities By Notional Buckets - 25MM+}
\end{figure}

\begin{doublespace}
As secondary evidence of increased costs, we present daily 90-day
moving volatilities on all three of trading costs metrics in Figures
\ref{Trading Costs and Volatilities}, \ref{Trading Cost Volatilities By Notional Buckets - 10MM+}
and \ref{Trading Cost Volatilities By Notional Buckets - 25MM+}.
The higher efforts in finding liquidity are seen from the Market Impact
volatilities which has risen consistently since the changes. We need
to keep in mind that the metric we are using is a moving 90 day volatility
hence the actual effects of the change start to show up after a few
days time. Also, the values near the start of the sample are not yet
fully incorporating many days of data and hence need to be overlooked.
The Market Timing and Total Slippage can vary from positive to negative
numbers, hence to calculate the corresponding volatilities we cannot
use continuous compounding and instead we use a 5 day moving average
and the percentage difference between successive values. Because these
values can fluctuate more widely than the Market Impact, the consistent
increasing volatility pattern is not easily inferred for these two
variables. Further explorations using intraday data are required to
establish whether higher volatility levels have been reached and could
be one possible explanation for higher trading costs.

In short, we have decreased spreads, decreased trading size, increased
number of trades, increased volume and increased trading costs for
larger orders. The implications of this and to whom the immediate
benefits will accrue should be fairly obvious. Buying and selling
smaller sizes more frequently can be less expensive, but buying (or
selling) and holding (after holding) larger chunks of securities for
longer periods of time might have become more costly. The aftereffects
of the change are not exactly a win-win situation for everyone, there
seem to be some winners and some losers.
\end{doublespace}
\begin{doublespace}

\section{Possibilities for a Deeper Dive}
\end{doublespace}

\begin{doublespace}
A key metric that would be useful to understand the effects of the
change would be the intraday volatilities of the prices and execution
sizes. Volatility is proportional to trading costs, hence, measuring
intraday volatility before and after the change could provide some
answers to why costs have increased. When price volatility and execution
size volatility increase, a trader faces a more uncertain environment.
He has to factor in his decisions the possibility of the price and
liquidity slipping away from him which results in higher overall costs
or uncertainty about costs, which is also costly. This happens through
greater swings in the Market Impact; in addition, greater movements
in the Market Timing will cause higher overall Slippage numbers. Intraday
data will allow the depth posted at the quotes to be analyzed and
this could explain the reduction in the execution sizes and the increased
difficulty in sourcing large orders. It would be interesting to trace
the number of cancelled orders, quotes, the additional messages being
relayed, changes to instructions and other forms of noise, and the
technology infrastructure being deployed to accommodate any additional
processing burdens, both by the exchange and (if possible to estimate)
across other participants. 
\end{doublespace}

Another intended effect of this event was to reduce the time to execution
which can be measured using tick by tick data. Again, it is worth
pondering the reasons why filling an order in 10 milliseconds or 50
milliseconds would make such a difference to the loftier goal of providing
a secondary market for the transfer of firm ownership and risk. For
the deep dive, we have not considered the results by the securities
affected during each event, since the results seem to hold strongly
across the entire set. Checking this extra box might show other potentially
interesting or unexpected outcomes. The sudden spread increase and
surrounding market events on Oct-31-2014 are worthy of a closer inspection.
This study has been performed as a completely closed system. To further
this avenue of approach, having external control variables could help
account for some of the cyclical or structural variations in the variables
and establish the trends strongly. This is particularly important
for the trading costs which are on a decreasing trend from the beginning
of our sample and including explanatory variables could explain this
phenomenon. We have not included moving trading cost volatilities
in any of our regressions, but these could be useful control variables
since these change at a slower pace compared to the actual variable
and pick up long term trends.
\begin{doublespace}

\section{Does Tick Size Matter? Tick Size Does Matter! }
\end{doublespace}

\begin{doublespace}
We conclude that one set of changes have happened as anticipated,
with the reduction in the average spread size. The unanticipated change
is the reduction in the average execution size. We also see that the
total volume and the number of trades have increased. The increased
volume is not necessarily proving beneficial in adding liquidity to
all exchange participants. On one hand, investors who are trading
smaller order sizes are likely to experience a decreased cost of trading;
on the other hand, large institutional investors trading bigger orders,
might require additional efforts to source in the liquidity to fill
their trades; with the net effect being that this additional effort
might even lead to a slightly increased effective cost of trading.
Once the dust from the last set of changes, which are yet to be implemented,
settles down, supplementing this study with more intraday indicators
will go a long way towards determining conclusively which group of
investors will be the ultimate beneficiary. In our attempt to answer
the question, ``Does Tick Size Matter?'', we unequivocally find
that, ``Tick Size Does Matter''. The significant competition between
trading mechanisms and venues, highlights the need for future research
related to the consequences of tick size on trading costs and the
dynamics of liquidity supply. 
\end{doublespace}
\begin{doublespace}

\section{Acknowledgements and End-notes}
\end{doublespace}
\begin{enumerate}
\begin{doublespace}
\item The following individuals have been a constant source of inputs and
encouragement, more continuous than the flow of orders in an extremely
liquid venue: Brad Hunt, Henry Yegerman, Samuel Zou, Alex Gillula
and Ronald Ang at Markit; Dr. Isabel Yan, Dr. Yong Wang, Dr. Vikas
Kakkar, Dr. Fred Kwan, Dr. Costel Daniel Andonie and Dr. Humphrey
Tung at the City University of Hong Kong. The views and opinions expressed
in this article, along with any mistakes, are mine alone and do not
necessarily reflect the official policy or position of either of my
affiliations or any other agency.
\item \label{enu:Sergey-Nazarovich-Bubka}Sergey Nazarovich Bubka (born
4 December 1963) is a Ukrainian former pole vaulter. He represented
the Soviet Union until its dissolution in 1991. Sergey has also beaten
his own record 14 times. He was the first pole vaulter to clear 6.0
metres and 6.10 metres. Bubka was twice named Athlete of the Year
by Track \& Field News and in 2012 was one of 24 athletes inducted
as inaugural members of the International Association of Athletics
Federations Hall of Fame. \href{https://en.wikipedia.org/wiki/Sergey_Bubka}{Sergey Bubka, Wikipedia Link}
\item \label{enu:Japanese-stocks-soared,}Japanese stocks soared, with the
Nikkei 225 Stock Average closing at a seven-year high, as the Bank
of Japan unexpectedly boosted easing and the nation\textquoteright s
pension fund prepared to unveil new asset allocations. \href{http://www.bloomberg.com/news/articles/2014-10-31/japan-stocks-rise-on-report-pension-fund-to-boost-shares}{Japan\textquoteright s Nikkei 225 Soars to Seven-Year High on BOJ, GPIF}
\item \label{enu:Japanese-shares-swing}Japanese shares are swinging by
the most on record after a double boost by the nation\textquoteright s
central bank and pension fund sent the Topix index to a six-year high
two weeks after it entered a correction. \href{http://www.bloomberg.com/news/articles/2014-11-05/sell-buy-sell-again-in-27-days-amid-record-topix-swings}{Sell, Buy, Sell Again in 27 Days Amid Record Topix Swings}
\end{doublespace}
\item \label{enu:Endogeneity-is-a}Endogeneity is a symptom in regression
models which causes the coefficients of the explanatory variables
to be biased and without additional information the coefficients cannot
be consistently estimated. Endogeneity arises when the error term
and the explanatory variables are correlated. The most common workaround
for endogeneity is to use an additional variable that is uncorrelated
with the error term but is correlated with the explanatory variable
in the original model. This technique for rectifying endogeneity is
known as instrumental variables regression method. \href{https://en.wikipedia.org/wiki/Endogeneity_(econometrics)}{Endogeneity (econometrics), Wikipedia Link}
\end{enumerate}

\section{References}
\begin{enumerate}
\begin{doublespace}
\item Ahn, H. J., Cao, C. Q., \& Choe, H. (1998). Decimalization and competition
among stock markets: Evidence from the Toronto Stock Exchange cross-listed
securities. Journal of Financial Markets, 1(1), 51-87.
\item Ahn, H. J., Cai, J., Chan, K., \& Hamao, Y. (2007). Tick size change
and liquidity provision on the Tokyo Stock Exchange. Journal of the
Japanese and International Economies, 21(2), 173-194.
\item Almgren, R., \& Chriss, N. (2001). Optimal execution of portfolio
transactions. Journal of Risk, 3, 5-40.
\item Almgren, R. F. (2003). Optimal execution with nonlinear impact functions
and trading-enhanced risk. Applied mathematical finance, 10(1), 1-18.
\item Almgren, R., Thum, C., Hauptmann, E., \& Li, H. (2005). Direct estimation
of equity market impact. Risk, 18, 5752.
\item Bessembinder, H. (1999). Trade execution costs on Nasdaq and the NYSE:
A post-reform comparison. Journal of Financial and Quantitative Analysis,
34(3).
\item Bessembinder, H. (2000). Tick size, spreads, and liquidity: An analysis
of Nasdaq securities trading near ten dollars. Journal of Financial
Intermediation, 9(3), 213-239.
\item Bessembinder, H. (2003). Trade execution costs and market quality
after decimalization. Journal of Financial and Quantitative Analysis,
38(04), 747-777.
\item Bhide, A. (1993). The hidden costs of stock market liquidity. Journal
of financial economics, 34(1), 31-51.
\item Bollen, N. P., \& Busse, J. A. (2006). Tick size and institutional
trading costs: Evidence from mutual funds. Journal of Financial and
Quantitative Analysis, 41(04), 915-937.
\item Bourghelle, D., \& Declerck, F. (2004). Why markets should not necessarily
reduce the tick size. Journal of banking \& finance, 28(2), 373-398.
\item Caldarelli, G., Marsili, M., \& Zhang, Y. C. (1997). A prototype model
of stock exchange. EPL (Europhysics Letters), 40(5), 479.
\item Chung, K. H., \& Van Ness, R. A. (2001). Order handling rules, tick
size, and the intraday pattern of bid\textendash ask spreads for Nasdaq
stocks. Journal of Financial Markets, 4(2), 143-161.
\item Collins, B. M., \& Fabozzi, F. J. (1991). A methodology for measuring
transaction costs. Financial Analysts Journal, 47(2), 27-36.
\item Easley, D., \& O'Hara, M. (2010). Microstructure and Ambiguity. Journal
of Finance, 65(5), 1817-1846.
\item Goldstein, M. A., \& Kavajecz, K. A. (2000). Eighths, sixteenths,
and market depth: changes in tick size and liquidity provision on
the NYSE. Journal of Financial Economics, 56(1), 125-149.
\item Gujarati, D. N. (1995). Basic econometrics, 3rd. International Edition.
\item Hamilton, J. D. (1994). Time series analysis (Vol. 2). Princeton university
press.
\item Harris, L. E. (1994). Minimum price variations, discrete bid-ask spreads,
and quotation sizes. Review of Financial Studies, 7(1), 149-178.
\item Huang, R. D., \& Stoll, H. R. (2001). Tick size, bid-ask spreads,
and market structure. Journal of Financial and Quantitative Analysis,
36(04), 503-522.
\item Jones, C. M., \& Lipson, M. L. (2001). Sixteenths: direct evidence
on institutional execution costs. Journal of Financial Economics,
59(2), 253-278.
\item Michie, R. (2001). The London stock exchange: A history. OUP Catalogue.
\item Karolyi, G. A., \& Stulz, R. M. (1996). Why do markets move together?
An investigation of US-Japan stock return comovements. The Journal
of Finance, 51(3), 951-986.
\item Kashyap, R. (2014a). Dynamic Multi-Factor Bid-Offer Adjustment Model.
Institutional Investor Journals, Journal of Trading, 9(3), 42-55.
\item Kashyap, R. (2014b). The Circle of Investment. International Journal
of Economics and Finance, 6(5), 244-263.
\item Kashyap, R. (2015a). Financial Services, Economic Growth and Well-Being:
A Four Pronged Study. Indian Journal of Finance, 9(1), 9-22.
\item Kashyap, R. (2015b). Hong Kong - Shanghai Connect / Hong Kong - Beijing
Disconnect (?). Social Science Research Network (SSRN), Working Paper.
\item Kim, K. A., \& Rhee, S. (1997). Price limit performance: evidence
from the Tokyo Stock Exchange. The Journal of Finance, 52(2), 885-901.
\item Kissell, R. (2006). The expanded implementation shortfall: Understanding
transaction cost components. The Journal of Trading, 1(3), 6-16.
\item Madhavan, A. (2000). Market microstructure: A survey. Journal of financial
markets, 3(3), 205-258.
\item Perold, A. F. (1988). The implementation shortfall: Paper versus reality.
The Journal of Portfolio Management, 14(3), 4-9.
\item Roll, R. (1984). A simple implicit measure of the effective bid-ask
spread in an efficient market. The Journal of Finance, 39(4), 1127-1139.
\item Stoll, H. R. (1989). Inferring the components of the bid-ask spread:
theory and empirical tests. The Journal of Finance, 44(1), 115-134.
\item Treynor, J. L. (1981). What does it take to win the trading game?.
Financial Analysts Journal, 37(1), 55-60.
\item Treynor, J. L. (1994). The invisible costs of trading. The Journal
of Portfolio Management, 21(1), 71-78.
\item Venkataraman, K. (2001). Automated versus floor trading: An analysis
of execution costs on the Paris and New York exchanges. The Journal
of Finance, 56(4), 1445-1485.
\item Ye, M., Yao, C., \& Gai, J. (2012). The externalities of high frequency
trading. SSRN Working Paper: http://papers.ssrn.com/abstract\_id=2066839
\item Yegerman, H. \& Gillula, A. (2014). The Use and Abuse of Implementation
Shortfall. Markit Working Paper.
\end{doublespace}
\end{enumerate}
\begin{doublespace}

\section{Appendix - I (Bird's Eye View Comparisons)\label{sec:Appendix---I}}
\end{doublespace}

\begin{doublespace}

\begin{figure}[H]
\includegraphics[width=18cm,height=23.8cm]{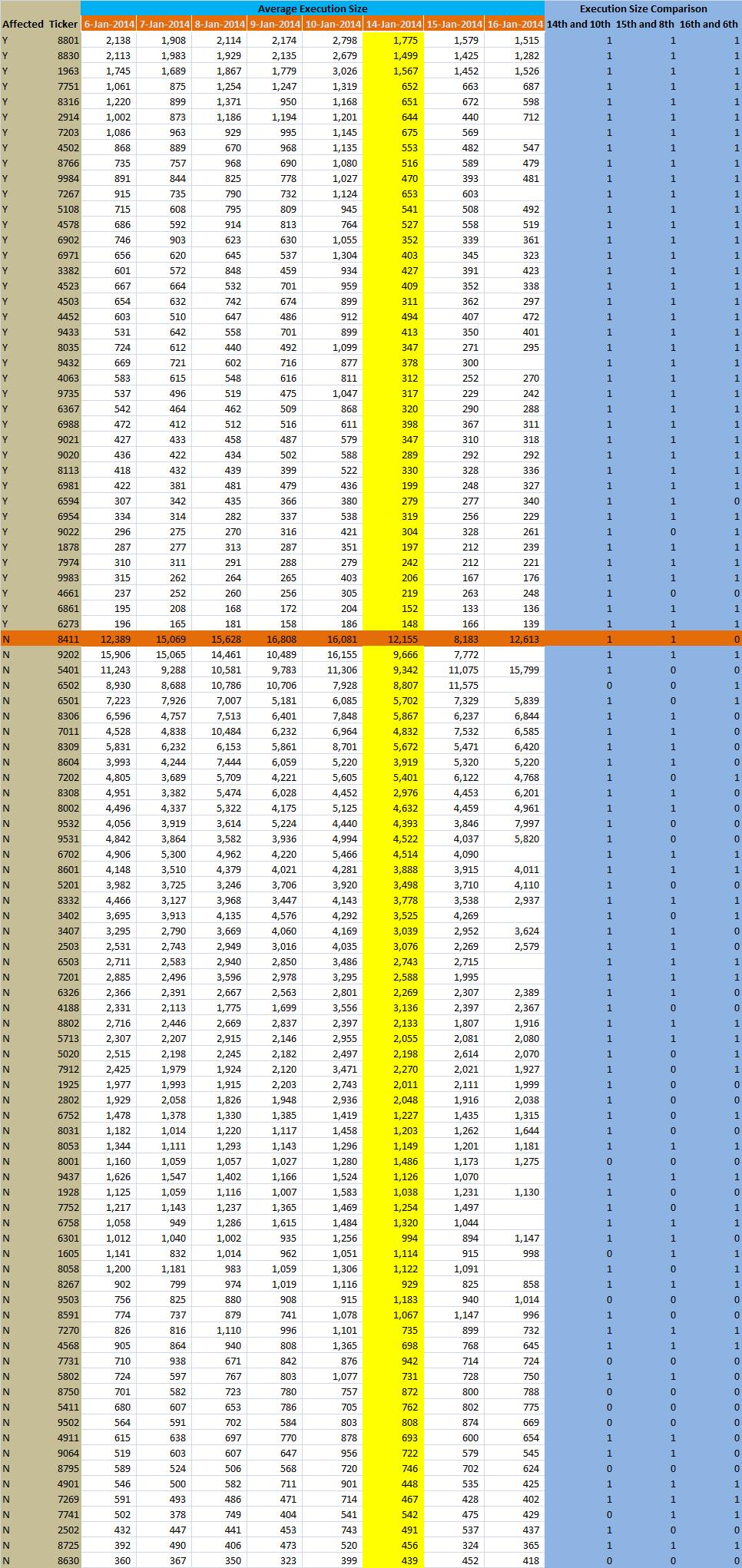}

\caption{Bird's Eye View Execution Size Comparison Detail for Jan 14, 2014}

\end{figure}

\begin{figure}[H]
\includegraphics[width=18cm,height=23.8cm]{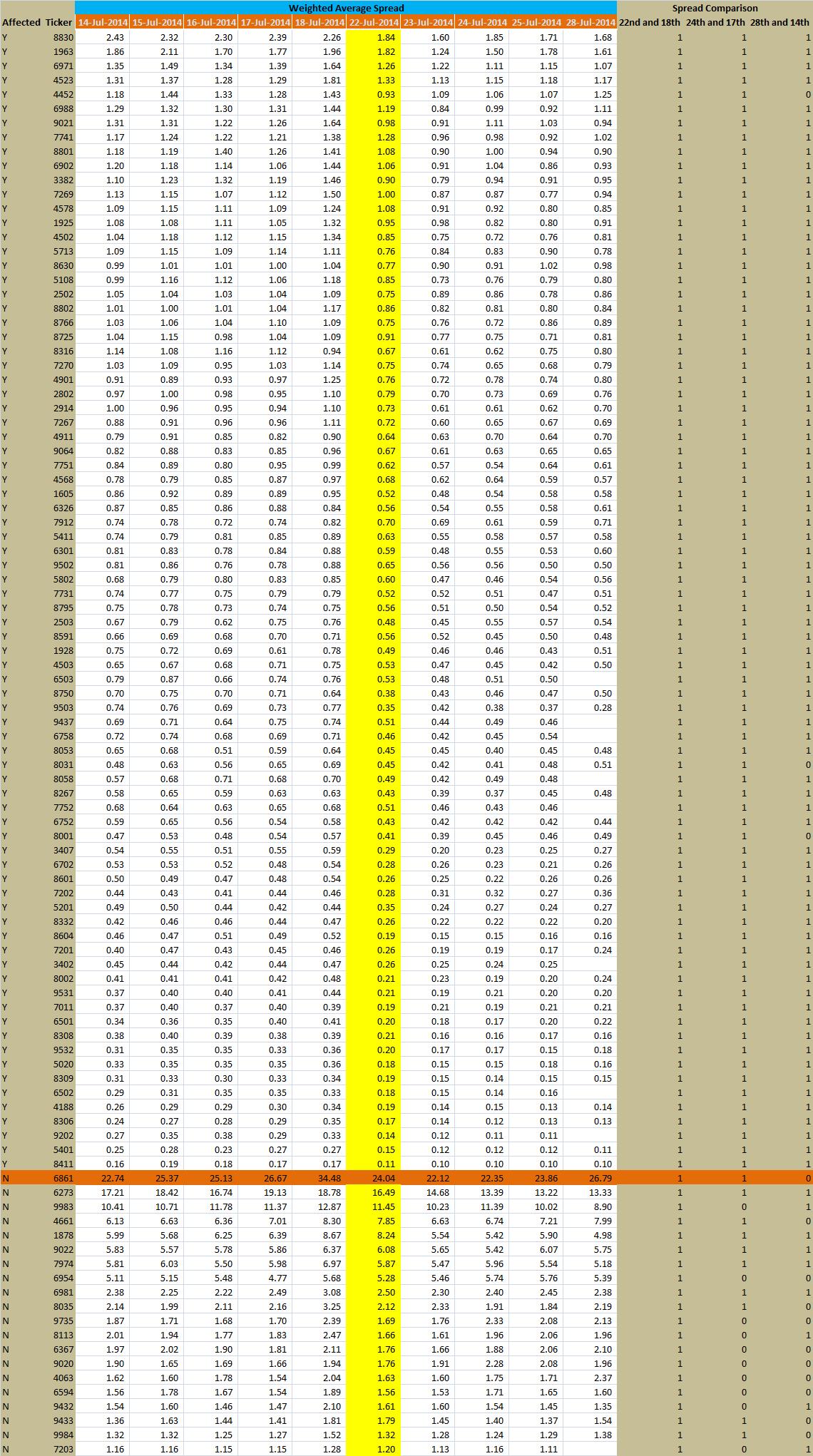}

\caption{Bird's Eye View Spread Comparison Detail for Jul 22, 2014}

\end{figure}

\end{doublespace}
\begin{doublespace}

\section{Appendix - II (Deep Dive Comparisons)\label{sec:Appendix---II}}
\end{doublespace}
\begin{itemize}
\begin{doublespace}
\item EW stands for Equal Weighted; VW for Volume Weighted; TW for Trade
Weighted and USDW for USD Weighted.
\end{doublespace}
\end{itemize}
\begin{doublespace}

\end{doublespace}


\section{Appendix - III (Proof of Endogeneity Resolution)\label{sec:Appendix---III}}

Let us consider a multiple regression model, as shown below, with
$k$ explanatory variables $x_{i}$ with corresponding coefficients
$\beta_{i}$ ; $\beta_{0}$ is the intercept; $y$ is the dependent
variable and $u$ is the error term,
\[
y=\beta_{0}+\beta_{1}x_{1}+\cdots+\beta_{k}x_{k}+u
\]
If this is the true model, the following moment or orthogonality conditions
are satisfied,
\[
E\left(u\right)=0
\]
\[
E\left(x_{j}u\right)=0\quad;\quad j=1,\cdots,k
\]
Alternately these conditions are, 
\[
E\left(u\right)=0
\]
\begin{align*}
E\left(x_{1}u\right) & =0\\
\cdots\\
E\left(x_{k}u\right) & =0
\end{align*}
These can be rewritten as,
\[
E\left(y-\beta_{0}-\beta_{1}x_{1}-\cdots-\beta_{k}x_{k}\right)=0
\]
\begin{align*}
E\left(x_{1}\left[y-\beta_{0}-\beta_{1}x_{1}-\cdots-\beta_{k}x_{k}\right]\right) & =0\\
\cdots\\
E\left(x_{k}\left[y-\beta_{0}-\beta_{1}x_{1}-\cdots-\beta_{k}x_{k}\right]\right) & =0
\end{align*}
This gives,
\[
E\left(y\right)=\beta_{0}+\beta_{1}E\left(x_{1}\right)+\cdots+\beta_{k}E\left(x_{k}\right)
\]
\begin{align*}
E\left(x_{1}y\right) & =\beta_{0}E\left(x_{1}\right)+\beta_{1}E\left(x_{1}^{2}\right)+\cdots+\beta_{k}E\left(x_{1}x_{k}\right)\\
\cdots\\
E\left(x_{k}y\right) & =\beta_{0}E\left(x_{k}\right)+\beta_{1}E\left(x_{k}x_{1}\right)+\cdots+\beta_{k}E\left(x_{k}^{2}\right)
\end{align*}
Writing this in matrix form,
\[
\left[\begin{array}{ccccc}
1 & E\left(x_{1}\right) & E\left(x_{2}\right) & \cdots & E\left(x_{k}\right)\\
E\left(x_{1}\right) & E\left(x_{1}^{2}\right) & E\left(x_{1}x_{2}\right) & \cdots & E\left(x_{1}x_{k}\right)\\
E\left(x_{2}\right) & E\left(x_{1}x_{2}\right) & E\left(x_{2}^{2}\right) & \cdots & E\left(x_{2}x_{k}\right)\\
\cdots & \cdots & \cdots & \cdots & \cdots\\
E\left(x_{k}\right) & E\left(x_{1}x_{k}\right) & E\left(x_{2}x_{k}\right) & \cdots & E\left(x_{k}^{2}\right)
\end{array}\right]\left[\begin{array}{c}
\beta_{0}\\
\beta_{1}\\
\beta_{2}\\
\cdots\\
\beta_{k}
\end{array}\right]=\left[\begin{array}{c}
E\left(y\right)\\
E\left(x_{1}y\right)\\
E\left(x_{2}y\right)\\
\cdots\\
E\left(x_{k}y\right)
\end{array}\right]
\]
This gives the solution or the coefficients as,
\[
\left[\begin{array}{c}
\beta_{0}\\
\beta_{1}\\
\beta_{2}\\
\cdots\\
\beta_{k}
\end{array}\right]=\left\{ \left[\begin{array}{ccccc}
1 & E\left(x_{1}\right) & E\left(x_{2}\right) & \cdots & E\left(x_{k}\right)\\
E\left(x_{1}\right) & E\left(x_{1}^{2}\right) & E\left(x_{1}x_{2}\right) & \cdots & E\left(x_{1}x_{k}\right)\\
E\left(x_{2}\right) & E\left(x_{1}x_{2}\right) & E\left(x_{2}^{2}\right) & \cdots & E\left(x_{2}x_{k}\right)\\
\cdots & \cdots & \cdots & \cdots & \cdots\\
E\left(x_{k}\right) & E\left(x_{1}x_{k}\right) & E\left(x_{2}x_{k}\right) & \cdots & E\left(x_{k}^{2}\right)
\end{array}\right]\right\} ^{-1}\left[\begin{array}{c}
E\left(y\right)\\
E\left(x_{1}y\right)\\
E\left(x_{2}y\right)\\
\cdots\\
E\left(x_{k}y\right)
\end{array}\right]
\]
In the presence of endogeneity or the case where the orthogonality
conditions are not satisfied, that is $E\left(x_{j}u\right)\neq0$,
we can write the correlation terms between each of the explanatory
variables and the error terms as $E\left(x_{j}u\right)=c_{j}$ or
$E\left(x_{j}u\right)-c_{j}=0$. This gives the coefficients in matrix
form as, 
\[
\left[\begin{array}{c}
\beta_{0}\\
\beta_{1}\\
\beta_{2}\\
\cdots\\
\beta_{k}
\end{array}\right]=\left\{ \left[\begin{array}{ccccc}
1 & E\left(x_{1}\right) & E\left(x_{2}\right) & \cdots & E\left(x_{k}\right)\\
E\left(x_{1}\right) & E\left(x_{1}^{2}\right) & E\left(x_{1}x_{2}\right) & \cdots & E\left(x_{1}x_{k}\right)\\
E\left(x_{2}\right) & E\left(x_{1}x_{2}\right) & E\left(x_{2}^{2}\right) & \cdots & E\left(x_{2}x_{k}\right)\\
\cdots & \cdots & \cdots & \cdots & \cdots\\
E\left(x_{1}\right) & E\left(x_{1}x_{k}\right) & E\left(x_{2}x_{k}\right) & \cdots & E\left(x_{k}^{2}\right)
\end{array}\right]\right\} ^{-1}\left[\begin{array}{c}
E\left(y\right)\\
E\left(x_{1}y\right)-c_{1}\\
E\left(x_{2}y\right)-c_{2}\\
\cdots\\
E\left(x_{k}y\right)-c_{k}
\end{array}\right]
\]
We note that if we have a sample of $n$ observations we can write
the regression equations as,
\[
y_{i}=\beta_{0}+\beta_{1}x_{i1}+\cdots+\beta_{k}x_{ik}+u_{i}\quad;\;i=1,\cdots,n
\]
The sum of the squared residuals or error terms can be written as,
\[
\sum_{i=1}^{n}u_{i}^{2}=\sum_{i=1}^{i=n}\left(y_{i}-\beta_{0}-\sum_{j=1}^{k}\beta_{j}x_{j}\right)^{2}
\]
Alternately,
\begin{align*}
y_{1} & =\beta_{0}+\beta_{1}x_{11}+\cdots+\beta_{k}x_{1k}+u_{1}\\
 & \cdots\quad\cdots\quad\cdots\quad\cdots\\
y_{n} & =\beta_{0}+\beta_{1}x_{n1}+\cdots+\beta_{k}x_{nk}+u_{n}
\end{align*}
Writing this in matrix form,
\[
\left[\begin{array}{ccccc}
1 & x_{11} & x_{12} & \cdots & x_{1k}\\
1 & x_{21} & x_{22} & \cdots & x_{2k}\\
1 & x_{31} & x_{32} & \cdots & x_{3k}\\
\cdots & \cdots & \cdots & \cdots & \cdots\\
1 & x_{n1} & x_{n2} & \cdots & x_{nk}
\end{array}\right]\left[\begin{array}{c}
\beta_{0}\\
\beta_{1}\\
\beta_{2}\\
\cdots\\
\beta_{k}
\end{array}\right]+\left[\begin{array}{c}
u_{1}\\
u_{2}\\
u_{3}\\
\cdots\\
u_{n}
\end{array}\right]=\left[\begin{array}{c}
y_{1}\\
y_{2}\\
y_{3}\\
\cdots\\
y_{n}
\end{array}\right]
\]
\[
\boldsymbol{y}=\boldsymbol{X\beta}+\boldsymbol{u}
\]
Here bold letters in lowercase denote vectors and bold letters in
uppercase denote matrices,
\[
\left[\begin{array}{c}
u_{1}\\
u_{2}\\
u_{3}\\
\cdots\\
u_{n}
\end{array}\right]=\left[\begin{array}{c}
y_{1}\\
y_{2}\\
y_{3}\\
\cdots\\
y_{n}
\end{array}\right]-\left[\begin{array}{ccccc}
1 & x_{11} & x_{12} & \cdots & x_{1k}\\
1 & x_{21} & x_{22} & \cdots & x_{2k}\\
1 & x_{31} & x_{32} & \cdots & x_{3k}\\
\cdots & \cdots & \cdots & \cdots & \cdots\\
1 & x_{n1} & x_{n2} & \cdots & x_{nk}
\end{array}\right]\left[\begin{array}{c}
\beta_{0}\\
\beta_{1}\\
\beta_{2}\\
\cdots\\
\beta_{k}
\end{array}\right]
\]
\[
\boldsymbol{u}=\boldsymbol{y}-\boldsymbol{X\beta}
\]
\[
\sum_{i=1}^{n}u_{i}^{2}=\boldsymbol{u^{T}}\boldsymbol{u}
\]
\[
=\boldsymbol{\left(\boldsymbol{y}-\boldsymbol{X\beta}\right)^{T}}\left(\boldsymbol{y}-\boldsymbol{X\beta}\right)
\]
The smallest the sum of squares of the error term could be is zero.
This is the case when all the $u_{i}$ are zero. Let $\boldsymbol{\hat{\beta}}$
denote the coefficients that minimize the mean squared error based
on the available sample, which gives the fitted equation for the regression
model or the predicted dependent values as follows,
\[
\boldsymbol{\hat{y}}=\boldsymbol{X\hat{\beta}}
\]
Here $\boldsymbol{\hat{y}}$ is the projection of the $n$-dimensional
data vector $\boldsymbol{y}$ onto the hyperplane spanned by $\boldsymbol{X}$,
\[
\boldsymbol{u}=\boldsymbol{y}-\boldsymbol{\hat{y}}
\]
\[
\boldsymbol{u}=\boldsymbol{y}-\boldsymbol{X\hat{\beta}}
\]
Using the orthogonality condition,
\[
\boldsymbol{X^{T}u}=0
\]
\[
\boldsymbol{\left(X^{T}\right)\left(\boldsymbol{y}-\boldsymbol{X\hat{\beta}}\right)}=0
\]
\[
\boldsymbol{\left(X^{T}\boldsymbol{y}\right)-X^{T}\boldsymbol{X\hat{\beta}}}=0
\]
\[
\boldsymbol{\boldsymbol{\hat{\beta}}=\left(X^{T}X\right)^{-1}\left(X^{T}\boldsymbol{y}\right)}
\]
\[
\boldsymbol{\left(X^{T}X\right)}=\left[\begin{array}{ccccc}
1 & 1 & 1 & \cdots & 1\\
x_{11} & x_{21} & x_{31} & \cdots & x_{n1}\\
x_{12} & x_{22} & x_{32} & \cdots & x_{n2}\\
\cdots & \cdots & \cdots & \cdots & \cdots\\
x_{1k} & x_{2k} & x_{3k} & \cdots & x_{nk}
\end{array}\right]\left[\begin{array}{ccccc}
1 & x_{11} & x_{12} & \cdots & x_{1k}\\
1 & x_{21} & x_{22} & \cdots & x_{2k}\\
1 & x_{31} & x_{32} & \cdots & x_{3k}\\
\cdots & \cdots & \cdots & \cdots & \cdots\\
1 & x_{n1} & x_{n2} & \cdots & x_{nk}
\end{array}\right]
\]
\begin{align*}
\boldsymbol{\left(X^{T}X\right)} & =\left[\begin{array}{cc}
1+1+\cdots+1 & x_{11}+x_{21}+x_{31}+\cdots+x_{n1}\\
x_{11}+x_{21}+x_{31}+\cdots+x_{n1} & x_{11}^{2}+x_{21}^{2}+x_{31}^{2}+\cdots+x_{n1}^{2}\\
x_{12}+x_{22}+x_{32}+\cdots+x_{n2} & x_{11}x_{12}+x_{21}x_{22}+x_{31}x_{32}+\cdots+x_{n1}x_{n2}\\
\cdots & \cdots\\
x_{1k}+x_{2k}+x_{3k}+\cdots+x_{nk} & x_{11}x_{1k}+x_{21}x_{2k}+x_{31}x_{3k}+\cdots+x_{n1}x_{nk}
\end{array}\right.\\
 & \left.\begin{array}{ccc}
x_{12}+x_{22}+x_{32}+\cdots+x_{n2} & \cdots & x_{1k}+x_{2k}+x_{3k}+\cdots+x_{nk}\\
x_{11}x_{12}+x_{21}x_{22}+x_{31}x_{32}+\cdots+x_{n1}x_{n2} & \cdots & x_{11}x_{1k}+x_{21}x_{2k}+x_{31}x_{3k}+\cdots+x_{n1}x_{nk}\\
x_{12}^{2}+x_{22}^{2}+x_{32}^{2}+\cdots+x_{n2}^{2} & \cdots & x_{12}x_{1k}+x_{22}x_{2k}+x_{32}x_{3k}+\cdots+x_{n2}x_{nk}\\
\cdots & \cdots & \cdots\\
x_{12}x_{1k}+x_{22}x_{2k}+x_{32}x_{3k}+\cdots+x_{n2}x_{nk} & \cdots & x_{1k}^{2}+x_{2k}^{2}+x_{3k}^{2}+\cdots+x_{nk}^{2}
\end{array}\right]
\end{align*}
Using the law of large numbers the above sample moments tend to the
population moments in probability and can be written as,
\[
\boldsymbol{\left(X^{T}X\right)}\overset{p}{\longrightarrow}n\left[\begin{array}{ccccc}
1 & E\left(x_{1}\right) & E\left(x_{2}\right) & \cdots & E\left(x_{k}\right)\\
E\left(x_{1}\right) & E\left(x_{1}^{2}\right) & E\left(x_{1}x_{2}\right) & \cdots & E\left(x_{1}x_{k}\right)\\
E\left(x_{2}\right) & E\left(x_{1}x_{2}\right) & E\left(x_{2}^{2}\right) & \cdots & E\left(x_{2}x_{k}\right)\\
\cdots & \cdots & \cdots & \cdots & \cdots\\
E\left(x_{k}\right) & E\left(x_{1}x_{k}\right) & E\left(x_{2}x_{k}\right) & \cdots & E\left(x_{k}^{2}\right)
\end{array}\right]
\]
Similarly,
\[
\left(\boldsymbol{X^{T}}\boldsymbol{y}\right)=\left[\begin{array}{ccccc}
1 & 1 & 1 & \cdots & 1\\
x_{11} & x_{21} & x_{31} & \cdots & x_{n1}\\
x_{12} & x_{22} & x_{32} & \cdots & x_{n2}\\
\cdots & \cdots & \cdots & \cdots & \cdots\\
x_{1k} & x_{2k} & x_{3k} & \cdots & x_{nk}
\end{array}\right]\left[\begin{array}{c}
y_{1}\\
y_{2}\\
y_{3}\\
\cdots\\
y_{n}
\end{array}\right]
\]
\[
\left(\boldsymbol{X^{T}}\boldsymbol{y}\right)=\left[\begin{array}{c}
y_{1}+y_{2}+y_{3}+\cdots+y_{k}\\
x_{11}y_{1}+x_{21}y_{2}+x_{31}y_{3}+\cdots x_{n1}y_{n}\\
x_{12}y_{1}+x_{22}y_{2}+x_{32}y_{3}+\cdots x_{n2}y_{n}\\
\cdots\\
x_{1k}y_{1}+x_{2k}y_{2}+x_{3k}y_{3}+\cdots x_{nk}y_{n}
\end{array}\right]
\]
\[
\left(\boldsymbol{X^{T}}\boldsymbol{y}\right)\overset{p}{\longrightarrow}n\left[\begin{array}{c}
E\left(y\right)\\
E\left(x_{1}y\right)\\
E\left(x_{2}y\right)\\
\cdots\\
E\left(x_{k}y\right)
\end{array}\right]
\]
Likewise, 
\[
\left(\boldsymbol{X^{T}}\boldsymbol{u}\right)=\left[\begin{array}{ccccc}
1 & 1 & 1 & \cdots & 1\\
x_{11} & x_{21} & x_{31} & \cdots & x_{n1}\\
x_{12} & x_{22} & x_{32} & \cdots & x_{n2}\\
\cdots & \cdots & \cdots & \cdots & \cdots\\
x_{1k} & x_{2k} & x_{3k} & \cdots & x_{nk}
\end{array}\right]\left[\begin{array}{c}
u_{1}\\
u_{2}\\
u_{3}\\
\cdots\\
u_{n}
\end{array}\right]
\]
\[
\left(\boldsymbol{X^{T}}\boldsymbol{u}\right)=\left[\begin{array}{c}
u_{1}+u_{2}+u_{3}+\cdots+u_{k}\\
x_{11}u_{1}+x_{21}u_{2}+x_{31}u_{3}+\cdots x_{n1}u_{n}\\
x_{12}u_{1}+x_{22}u_{2}+x_{32}u_{3}+\cdots x_{n2}u_{n}\\
\cdots\\
x_{1k}u_{1}+x_{2k}u_{2}+x_{3k}u_{3}+\cdots x_{nk}u_{n}
\end{array}\right]
\]
\[
\left(\boldsymbol{X^{T}}\boldsymbol{u}\right)\overset{p}{\longrightarrow}n\left[\begin{array}{c}
E\left(u\right)\\
E\left(x_{1}u\right)\\
E\left(x_{2}u\right)\\
\cdots\\
E\left(x_{k}u\right)
\end{array}\right]
\]
In the presence of endogeneity the orthogonality condition, $\boldsymbol{X^{T}u}=0$,
does not hold and has to be modified as,
\[
\boldsymbol{X^{T}u}=\boldsymbol{c}
\]
Here $\boldsymbol{c}$ is a vector that holds the covariance of the
error term with each of the explanatory variables, that is $c_{j}=E\left(x_{j}u\right)$.
Noting that $E\left(u\right)=0$ gives,
\[
\boldsymbol{c}=\left[\begin{array}{c}
0\\
c_{1}\\
c_{2}\\
\cdots\\
c_{k}
\end{array}\right]
\]
\[
\boldsymbol{\left(X^{T}\right)\left(\boldsymbol{y}-\boldsymbol{X\hat{\beta}}\right)}=\boldsymbol{c}
\]
\[
\boldsymbol{\left(X^{T}\boldsymbol{y}\right)-X^{T}\boldsymbol{X\hat{\beta}}}=\boldsymbol{c}
\]
\[
\boldsymbol{\boldsymbol{\hat{\beta}}=\left(X^{T}X\right)^{-1}\left(X^{T}\boldsymbol{y}-\boldsymbol{c}\right)}
\]
\[
\boldsymbol{\boldsymbol{\hat{\beta}}=\left(X^{T}X\right)^{-1}\left(X^{T}\boldsymbol{y}\right)-\left(X^{T}X\right)^{-1}\boldsymbol{c}}
\]
This $\boldsymbol{\hat{\beta}}$, or the unbiased coefficients, cannot
be consistently estimated since $\boldsymbol{c}$, the covariance
of the error term with each of the explanatory variables, is unobserved.
What we end up calculating in the presence of endogeneity are the
coefficients, $\boldsymbol{\hat{\beta}_{E}}$ , given below,
\[
\boldsymbol{\hat{\beta}_{E}}=\boldsymbol{\left(X^{T}X\right)^{-1}\left(X^{T}\boldsymbol{y}\right)}
\]
\[
\boldsymbol{\hat{\beta}_{E}}=\boldsymbol{\boldsymbol{\hat{\beta}}+\left(X^{T}X\right)^{-1}\boldsymbol{c}}
\]
For any event study let the true coefficients before and after the
event be denoted by, $\boldsymbol{\hat{\beta}_{B}}$ and $\boldsymbol{\hat{\beta}_{A}}$.
Suppose we have not been able to completely eliminate endogeneity
the biased coefficients, $\boldsymbol{\hat{\beta}_{BE}}$ and $\boldsymbol{\hat{\beta}_{AE}}$,
are given by,
\[
\boldsymbol{\hat{\beta}_{BE}}=\boldsymbol{\boldsymbol{\hat{\beta}_{B}}+\left(X_{B}^{T}X_{B}\right)^{-1}\boldsymbol{c_{B}}}
\]
\[
\boldsymbol{\hat{\beta}_{AE}}=\boldsymbol{\boldsymbol{\hat{\beta}_{A}}+\left(X_{A}^{T}X_{A}\right)^{-1}\boldsymbol{c_{A}}}
\]
Here the suffix $\boldsymbol{B}$ and $\boldsymbol{A}$ denote values
before and after the event. The change in the coefficient values before
and after the event will be given by,
\[
\boldsymbol{\hat{\beta}_{AE}}-\boldsymbol{\hat{\beta}_{BE}}=\boldsymbol{\boldsymbol{\hat{\beta}_{A}}-\boldsymbol{\hat{\beta}_{B}}}\boldsymbol{+}\boldsymbol{\left(X_{A}^{T}X_{A}\right)^{-1}\boldsymbol{c_{A}}}\boldsymbol{-}\boldsymbol{\left(X_{B}^{T}X_{B}\right)^{-1}\boldsymbol{c_{B}}}
\]
Since we are considering only the change in the coefficients, we end
up measuring the true change in the coefficients as long as $\boldsymbol{\left(X_{B}^{T}X_{B}\right)^{-1}\boldsymbol{c_{B}}}\approx\boldsymbol{\left(X_{A}^{T}X_{A}\right)^{-1}\boldsymbol{c_{A}}}$.
What this suggests is that as long as the covariance structure among
the explanatory variables and the covariance between the error term
and the explanatory variables are comparable before and after the
event, we end up measuring the actual change between the coefficients.
This ensures that we are using the correct values to understand how
trading costs have altered before and after the event.
\end{document}